\documentclass{aastex631}

\begin{document}

\title{Multiband Optical Photometric and Spectroscopic Monitoring of the 2024 Flare Event in Transition Blazar OP313}

\author[0000-0002-7032-9667]{TianFang Zhang}
\affiliation{National Astronomical Observatory of Japan, National Institutes of Natural Science, 2-21-1 Osawa, Mitaka, Tokyo 181-8588, Japan}
\affiliation{Department of Astronomy, Graduate School of Science, The University of Tokyo, 7-3-1, Hongō, Bunkyo-ku, Tokyo 113-8654, Japan}
\affiliation{Institute of Astronomy, Graduate School of Science, The University of Tokyo, 2-21-1 Osawa, Mitaka, Tokyo 181-0015, Japan}

\author[0000-0001-6402-1415]{Mitsuru Kokubo}
\affiliation{National Astronomical Observatory of Japan, National Institutes of Natural Science, 2-21-1 Osawa, Mitaka, Tokyo 181-8588, Japan}

\author{Mamoru Doi}
\affiliation{National Astronomical Observatory of Japan, National Institutes of Natural Science, 2-21-1 Osawa, Mitaka, Tokyo 181-8588, Japan}
\affiliation{Institute of Astronomy, Graduate School of Science, The University of Tokyo, 2-21-1 Osawa, Mitaka, Tokyo 181-0015, Japan}
\affiliation{
Research center for the early universe School of Science Bldg. No.4, The University of Tokyo, 7-3-1, Hongō, Bunkyo-ku, Tokyo 113-0033, Japan}

\author{Haruna Hagio}
\affiliation{Department of Physics, Institute of Science Tokyo, 2-12-1 Ookayama, Meguro-ku, Tokyo 152-8550, Japan}

\author{Hibiki Seki}
\affiliation{Department of Physics, Institute of Science Tokyo, 2-12-1 Ookayama, Meguro-ku, Tokyo 152-8550, Japan}

\author[0000-0003-2691-4444]{Ichiro Takahashi}
\affiliation{Department of Physics, Institute of Science Tokyo, 2-12-1 Ookayama, Meguro-ku, Tokyo 152-8550, Japan}

\author[0000-0002-9579-731X]{Katsuhiro L. Murata}
\affiliation{Okayama Observatory, Kyoto University, 3037-5 Honjo, Kamogatacho, Asakuchi, Okayama 719-0232, Japan}

\author[0000-0001-6473-5100]{Kazuya Matsubayashi}
\affiliation{Institute of Astronomy, Graduate School of Science, The University of Tokyo, 2-21-1 Osawa, Mitaka, Tokyo 181-0015, Japan}

\author[0000-0002-6480-3799]{Keisuke Isogai}
\affiliation{Okayama Observatory, Kyoto University, 3037-5 Honjo, Kamogatacho, Asakuchi, Okayama 719-0232, Japan}
\affiliation{Department of Multi-Disciplinary Sciences, Graduate School of Arts and Sciences, The University of Tokyo, 3-8-1 Komaba, Meguro, Tokyo 153-8902, Japan}

\author[0000-0001-6099-9539]{Koji Kawabata}
\affiliation{Hiroshima Astrophysical Science Center, Hiroshima University, 1-3-1 Kagamiyama, Higashi-Hiroshima, Hiroshima 739-8526, Japan}

\author[0000-0001-5946-9960]{Mahito Sasada}
\affiliation{Institute of Integrated Research, Institute of Science Tokyo, 2-12-1 Ookayama, Meguro-ku, Tokyo 152-8550, Japan}

\author[0000-0003-3102-7452]{Masafumi Niwano}
\affiliation{Department of Physics, Institute of Science Tokyo, 2-12-1 Ookayama, Meguro-ku, Tokyo 152-8550, Japan}

\author[0009-0003-4534-9361]{Masaki Hashizume}
\affiliation{Physics Program, Graduate School of Advanced Science and Engineering, Hiroshima University, 1-3-1 Kagamiyama, Higashi-Hiroshima, Hiroshima 739-8526, Japan}

\author[0000-0001-8195-6546]{Megumi Shidatsu}
\affiliation{Department of Physics, Ehime University, 2-5, Bunkyocho, Matsuyama, Ehime 790-8577, Japan}

\author{Narikazu Higuchi}
\affiliation{Department of Physics, Institute of Science Tokyo, 2-12-1 Ookayama, Meguro-ku, Tokyo 152-8550, Japan}

\author[0000-0002-0643-7946]{Ryo Imazawa}
\affiliation{Physics Program, Graduate School of Advanced Science and Engineering, Hiroshima University, 1-3-1 Kagamiyama, Higashi-Hiroshima, Hiroshima 739-8526, Japan}

\author{Shigeaki Joshima}
\affiliation{Department of Physics, Institute of Science Tokyo, 2-12-1 Ookayama, Meguro-ku, Tokyo 152-8550, Japan}

\author[0000-0002-8792-2205]{Shigeyuki Sako}
\affiliation{Institute of Astronomy, Graduate School of Science, The University of Tokyo, 2-21-1 Osawa, Mitaka, Tokyo 181-0015, Japan}
\affiliation{UTokyo Organization for Planetary Space Science, The University of Tokyo, 7-3-1 Hongo, Bunkyo-ku, Tokyo 113-0033, Japan}
\affiliation{Collaborative Research Organization for Space Science and Technology, The University of Tokyo, 7-3-1 Hongo, Bunkyo-ku, Tokyo 113-0033, Japan}

\author{Shunsuke Hayatsu}
\affiliation{Department of Physics, Institute of Science Tokyo, 2-12-1 Ookayama, Meguro-ku, Tokyo 152-8550, Japan}

\author[0000-0003-1890-3913]{Yoichi Yatsu}
\affiliation{Department of Physics, Institute of Science Tokyo, 2-12-1 Ookayama, Meguro-ku, Tokyo 152-8550, Japan}

\author[0000-0002-0207-9010]{Wataru Iwakiri}
\affiliation{International Center for Hadron Astrophysics, Chiba University, Chiba 263-8522, Japan}

\author{Yoshiyuki Kubo}
\affiliation{Department of Physics, Institute of Science Tokyo, 2-12-1 Ookayama, Meguro-ku, Tokyo 152-8550, Japan}

\begin{abstract}
Blazars are active galactic nuclei known for their extreme variability, offering unique opportunities to study jet physics and high-energy emission mechanisms. In 2024, the Flat Spectrum Radio Quasar (FSRQ) OP313 underwent a remarkable flare event, during which the gamma-ray flux observed by the Fermi Large Area Telescope (Fermi/LAT) increased by a factor of 60 over its average value. The flare peak lasted less than two days. Using optical telescopes, we conducted 100-day time-scale observations. Multi-wavelength data revealed that OP313 entered an active state 50 days prior to the flare and remained active for at least 50 days afterward. We propose that this prolonged activity results from variations in electron density within the shock front due to changes in the accretion rate. Concurrently, OP313's spectrum transitioned from an FSRQ-like state to a BL Lac-like state, characterized by a significant increase in the synchrotron peak frequency and the disappearance of broad-line region emission lines. In the post-flare phase, we observed a decoupling between synchrotron radiation and inverse Compton scattering, along with a possible decrease in the magnetic field strength within the shock front. 

\end{abstract}


\clearpage
\clearpage
\newpage
\section{Introduction} \label{sec:intro}

Blazars represent one of the most extreme types of active galactic nuclei (AGNs) and are characterized by relativistic jets that are nearly aligned with our line of sight. This alignment leads to Doppler-boosted emission across the electromagnetic spectrum—from radio waves to gamma rays—and is responsible for their rapid flux variability and high polarization \citep{urry1995unified}. Gamma-ray observations are particularly valuable for probing the high-energy electron acceleration processes within these jets \citep{abdo2010spectral}, and allow us to quickly distinguish them from other AGN. Traditionally, blazars are divided into two subclasses: flat-spectrum radio quasars (FSRQs) and BL Lacertae objects (BL Lacs). FSRQs exhibit strong broad emission lines and are thought to host relatively luminous accretion disks, while BL Lacs display weak or absent broad lines, suggesting lower disk luminosities \citep{giommi2012simplified,ghisellini2011transition}. Radio-loud quasars were first identified in the early to mid-1960s (e.g., \citealt{schmidt19633c273}), leading to the discovery of optically violently variable (OVV) quasars in the 1970s \citep{basu1973optical}. The term “blazar” was introduced by Spiegel at a 1978 Pittsburgh conference \citep{BlandfordRees1978} and quickly became a unifying label for both OVV quasars and BL Lacs. Notably, BL Lac was originally cataloged as a variable star in 1929 \citep{hoffmeister1929354} and later recognized as an AGN in the 1970s \citep{miller1978spectrum}. By the mid-1980s, it was widely accepted that BL Lacs and FSRQs constitute two subclasses within the broader blazar family \citep{urry1995unified}.

Despite being grouped together as blazars, FSRQs and BL Lacs differ markedly in both their spectra and their host environments. FSRQs are high-ionization sources that reside in Fanaroff–Riley type II (FR II) radio galaxies—systems whose radio maps are edge-brightened and powered by strong, well-collimated jets ending in luminous hotspots. These objects exhibit prominent broad emission lines that signal the presence of a luminous accretion disk. By contrast, BL Lacs are low-ionization sources found in Fanaroff–Riley type I (FR I) radio galaxies, whose jets fade gradually with distance from the nucleus and lack bright terminal hotspots. Correspondingly, their optical spectra usually show weak or absent broad emission lines \citep{giommi2012simplified,fanaroff1974morphology}. Several criteria have been proposed to more robustly distinguish FSRQs from BL Lacs, including the equivalent widths (EW) of key broad lines (e.g., Mg II, H$\beta$), the spectral slope from optical to X-ray frequencies, the position of the synchrotron peak, and the inferred accretion rate relative to the Eddington limit \citep{ghisellini2011high}. In particular, an EW larger than 5 Å often signals a significantly ionizing disk, whereas sources with little or no detectable broad lines are classified as BL Lacs.

Recent observations challenge this long‐standing dichotomy. For example, \cite{ruan2014nature} showed that SDSS J083223.22+491321.0 (OJ448) changed its optical spectrum in just 13 days—evolving from a steep, BL Lac–like continuum to a flat, FSRQ–like shape without ever developing the broad emission lines typical of FSRQs. This rapid transition echoes the changing‐look quasar (CLQ) phenomenon, in which an AGN switches between Type 1 (broad lines present) and Type 2 (broad lines absent) states alongside dramatic luminosity swings (e.g., \citealt{ricci2023changing}). By the same token, other changing‐look blazars (CLBs) similar to OJ448 have now been identified: these objects move back and forth between the BL Lac and FSRQ classes as their flux and spectral properties vary substantially (e.g., \citealt{kang2024physical}). Additionally, \cite{pandey2024origin,pandey2024b2} presented multiwavelength observations of B2 1308+326 (OP313), showing that variations in its optical and $\gamma$-ray emissions can temporarily mimic a BL Lac state, even though its accretion disk and broad emission lines remain consistent with an FSRQ classification. Moreover, \cite{pena2021optical} used the LAMOST spectroscopic survey to determine new redshifts for blazar candidates, revealing other transitional systems in which optical lines and continuum states vary noticeably over short timescales. Similarly, \cite{graham2020understanding} investigated CLQs in the Catalina Real-time Transient Survey and observed sudden changes in both broad-line strengths and optical variability. Together, these studies suggest that some blazars can alternate between distinct spectral states. This transitional behavior implies that classifying blazars solely based on the presence or absence of emission lines may no longer be reliable. These findings highlight the need for more detailed monitoring of spectral transitions and high-cadence multiwavelength observations, as additional examples are necessary to clarify the underlying physical mechanisms.

New observational insights indicate potential strategies for identifying and monitoring such transitions. \cite{zhang2024optical} found that a subset of FSRQs with a gamma-ray photon index below 2.6 exhibits unusual optical variability. Notably, OJ448 itself has a gamma-ray photon index of 2.38 in the 4LAC catalog \citep{ajello2020fourth}, suggesting that FSRQs with similarly low photon indices may be prime candidates for detecting future transitions. By focusing on these sources, it may be possible to track and study the next event, thereby improving our understanding of blazar classification and the physics governing these remarkable objects.

\section{Spectroscopy and Multi-Band Observations of OP313}
\label{sec:Spectroscopy of OP313}

We initiated follow-up observations in response to Target of Opportunity (ToO) alerts, which were triggered by strong $\gamma$-ray variability detected by Fermi/LAT \citep{atwood2009large} and announced via The Astronomer’s Telegram (ATel)\citep{bartolini2024fermi}. On February 27, 2024, Fermi/LAT detected a sharp increase in the $\gamma$-ray flux from OP313 (4LAC gamma-ray photon index = 2.34; redshift = 1.00), rising to $3.1 \times 10^{-6}$ photons cm$^{-2}$ s$^{-1}$. This value was approximately 60 times the source’s previous daily average flux. Simultaneously, the measured $\gamma$-ray photon index decreased from 2.34, as recorded in the 4LAC catalog, to 1.8 $\pm$ 0.1. Motivated by this event, we began a spectroscopic monitoring using KOOLS-IFU starting on March 3, 2024.

\begin{figure}
    \centering
    \includegraphics[width=0.9\textwidth]{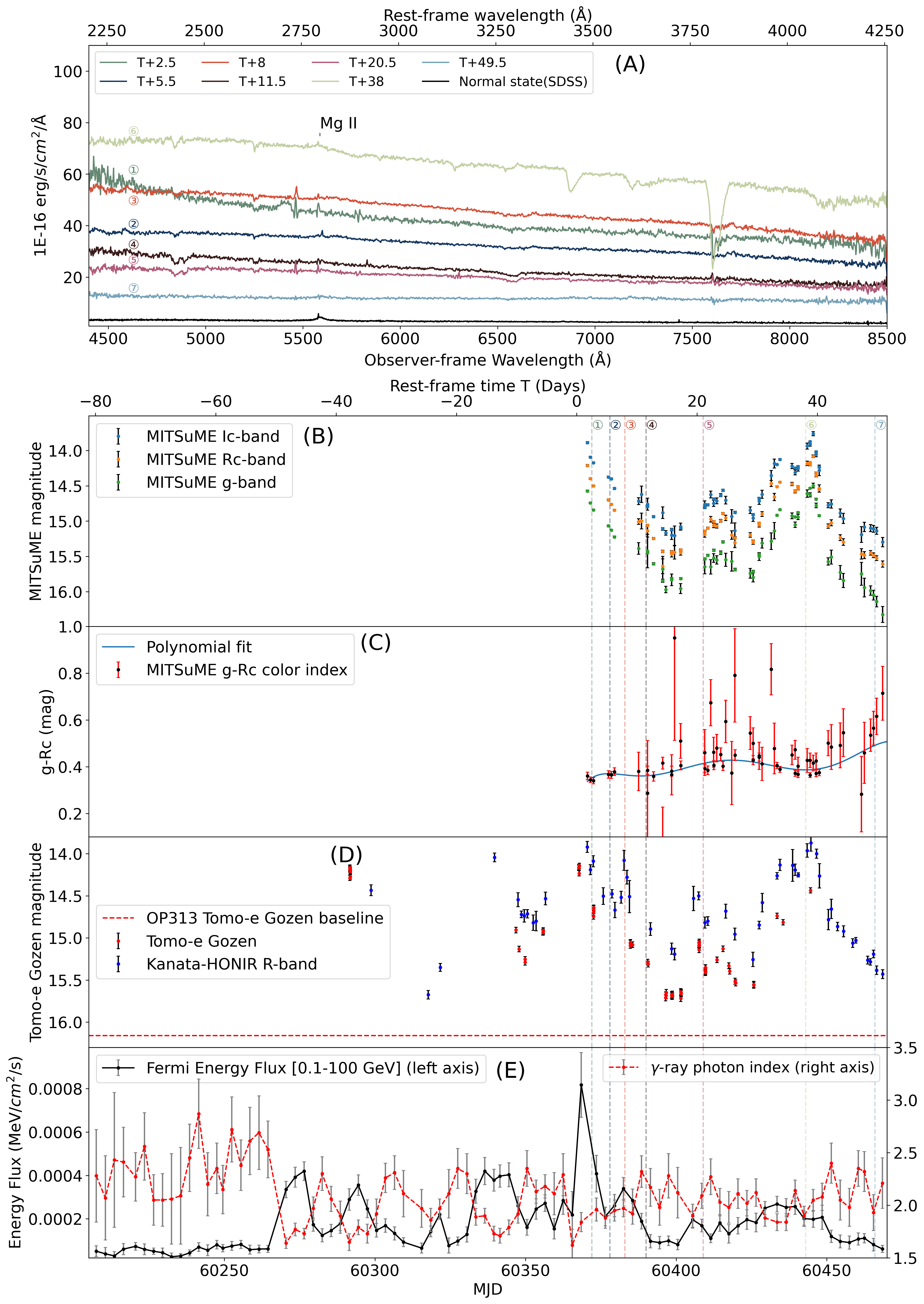}
    \caption{(A): The KOOLS-IFU and SDSS spectra of OP313, The position of the unresolved doublet of singly-ionized magnesium (Mg II 2796 \& 2803) is labeled; (B): The MITSuME 3-band light curve of OP313, error bars are omitted for clarity; (C): MITSuME g-Rc color index curve of of OP313; (D): Tomo-e Gozen (see \citep{zhang2024optical}) and HONIR light curve of OP313. The red dashed line represents the average magnitude of OP313 during the entire Tomo-e Gozen monitoring period; (E): Fermi/LAT energy flux (left axis) and $\gamma$-ray photon index (right axis) curve of OP313.}
    \label{fig:op313sep}
\end{figure}

Figure \ref{fig:op313sep} presents our multi-wavelength observations of OP313. Figure \ref{fig:op313sep} (A) shows seven spectra obtained with KOOLS-IFU on the Seimei 3.8m telescope \citep{matsubayashi2019kools} via Optical and Infrared Synergetic Telescopes for Education and Research (OISTER) program Time of Opportunity (ToO) observation, each with a typical integration time of 30 minutes using the VPH-blue grism, along with one archival Sloan Digital Sky Survey (SDSS) spectrum \citep{york2000sloan}. The KOOLS-IFU raw frames were processed with the public reduction pipeline developed by F. Iwamuro. The main steps are: (1) construction and subtraction of a master bias; (2) removal of cosmic-ray events; (3) creation of a pixel-to-pixel response map from twilight flats, followed by flat-field correction; (4) tracing of all 110 fibers, two-dimensional distortion mapping, and wavelength calibration using simultaneous Ne–Hg–Xe comparison-lamp exposures; (5) pairwise sky subtraction; (6) one-dimensional optimal extraction of each fiber spectrum and atmospheric-dispersion correction; (7) flux calibration using the spectrophotometric standard star; the pipeline automatically corrects for telluric absorption and applies a barycentric velocity correction. In Figure \ref{fig:op313sep} (A), T marks the time of the flare event, and T+X represents the number of days after the flare in the rest frame. The numbers within the circles indicate the chronological order of these observations. Atmospheric absorption features from oxygen (B-band) and water vapor are visible around 6880 Å and 7600 Å in the T+38 spectrum, resulting from imperfect standard-star correction. These features do not coincide with any emission lines from OP313’s broad-line region and thus do not affect our scientific analysis.

Figure \ref{fig:op313sep} (B) of Figure \ref{fig:op313sep} shows the three-color light curves obtained with the MITSuME camera at the AKENO and Okayama telescopes (\cite{kotani2007mitsume}, \cite{yatsu2007development}, \cite{shimokawabe2008mitsume}, \cite{yanagisawa2010six}) via OISTER program ToO observation (ID: 23B-K-0042).

Figure \ref{fig:op313sep} (C) shows the temporal evolution of the g-Rc color index from MITSuME. Although the presence of a dichroic mirror reduces the g-band transmittance below 4600 Å, limiting sensitivity to the blue-end brightening noted in the KOOLS-IFU data, polynomial fitting still reveals that the g-Rc color index increased from T+2.5 to T+5.5. After T+5.5, it continued to rise, albeit with fluctuations, throughout the remaining observation period.

Figure \ref{fig:op313sep} (D) presents the Tomo-e Gozen \citep{sako2016development} light curve on the Kiso 105 cm Schmidt telescope \citep{takase1977105} and the Hiroshima Optical and Near-InfraRed camera (HONIR \citep{akitaya2014honir}) R-band light curve on the 150 cm telescope, providing a record of OP313’s optical variability prior to the flare event. Finally, Figure \ref{fig:op313sep} (E) displays the Fermi/LAT energy flux (left axis) and the corresponding $\gamma$-ray photon index (right axis) before and after the $\gamma$-ray flare.

Table \ref{tab:instruments} summarizes the basic information of the instruments used in this observation. For MITSuME, HONIR, and Tomoe-Gozen, all observation details and raw data are publicly available through the Subaru-Mitaka-Okayama-Kiso Archive (SMOKA)\citep{baba2002development} operated by the National Astronomical Observatory of Japan. In the case of KOOLS-IFU, since the OISTER program used proprietary (non-OPEN use) observation time, the raw data are not accessible via SMOKA. Instead, the key observational information is summarized in Table \ref{tab:kools}.

\begin{table}[ht!]
  \caption{Basic parameters of observation instruments}
    \begin{center}
    \begin{tabular}{c|cccc}
    \hline
    Name & Telescope Aperture & Band & FoV & Limiting Magnitude  \\
    \hline
    KOOLS-IFU & 3.8m & 410$\sim$890 nm R=500 & 8.4*8.0 arcsec & 18.2$\sim$18.7 ABmag (30min exp S/N$\approx$5)\\
    MITSuME & 0.5m & 400$\sim$950 nm (g'-Rc-Ic)& 28*28 arcmin & g':16.7;Rc:16.6;Ic:15.8 (60sec exp, S/N$\approx$10)\\
    HONIR & 1.5m & 500$\sim$1000 nm & 15*15 arcmin & 18.9 ABmag (60sec exp S/N$\approx$10)\\
    Tomoe-Gozen & 1.05m & 300$\sim$1300 nm & 39.7*22.4*84 arcmin  & 17.7 ABmag (1sec exp S/N$\approx$10)\\
    \hline
    \end{tabular}%
    \end{center}
  \label{tab:instruments}%
\end{table}%

\begin{table}[ht!]
  \caption{KOOLS-IFU Observation Record}
    \begin{center}
    \begin{tabular}{c|cccc}
    \hline
    EXP No. & MJD-STR & EXP.TIME(sec) & ZD (secz) & S/N (@Mg II)  \\
    \hline
    1-1& 60372.5225 & 600 & 60.6134 (2.038) & 39.5 \\
    1-2& 60372.5297 & 312 & 58.6281 (1.921) & 26.7 \\
    1-3& 60372.5400 & 300 & 55.7369 (1.776) & 18.3 \\
    1-4& 60372.5437 & 300 & 54.7035 (1.731) & 29.1 \\
    1-5& 60372.5473 & 300 & 53.6674 (1.688) & 32.4 \\
    1-6& 60372.5510 & 300 & 52.6287 (1.648) & 35.8 \\
    \hline
    2-1& 60378.6946 & 600 & 5.8662 (1.005) & 91.2 \\
    2-2& 60378.7017 & 600 & 3.9978 (1.002) & 89.9 \\
    2-3& 60378.7088 & 600 & 2.6062 (1.001) & 83.1 \\
    \hline
    3-1& 60383.7654 & 600 & 20.1971 (1.066)	 & 104.2 \\
    3-2& 60383.7726 & 600 & 22.3199 (1.081)	 & 100.2 \\
    3-3& 60383.7797 & 600 & 24.4421 (1.098)  & 95.7 \\
    \hline
    4-1& 60390.5766 & 600 & 30.9202 (1.166)	 & 46.6 \\
    4-2& 60390.5838 & 600 & 28.8142 (1.141)	 & 45.0 \\
    4-3& 60390.5909 & 600 & 26.7041 (1.119)  & 52.0 \\
    \hline
    5-1& 60409.4701 & 600 & 46.8792 (1.463)	 & 44.9 \\
    5-2& 60409.4772 & 600 & 44.8183 (1.410)	 & 50.2 \\
    5-3& 60409.4843 & 600 & 42.7497 (1.362)  & 49.2 \\
    \hline
    6-1& 60443.6513 & 600 & 34.8963 (1.219)	 & 115.6 \\
    6-2& 60443.6585 & 600 & 36.9907 (1.252)	 & 114.4 \\
    7-3& 60443.6656 & 600 & 39.0782 (1.288)  & 111.7 \\
    \hline
    7-1& 60466.5242 & 840 & 15.8764 (1.040)	 & 35.8 \\
    7-2& 60466.5342 & 840 & 18.8203 (1.056)	 & 23.7 \\
    7-3& 60466.5441 & 900 & 21.7627 (1.077)  & 22.3 \\
    \hline
    \end{tabular}%
    \end{center}
  \label{tab:kools}%
\end{table}%

\clearpage
\section{Results}
\label{sec:Results}

We calculated the equivalent width (EW) of the Mg II 2798 Å line and its associated uncertainty for each spectrum separately, using the method as follows. First, each observed spectrum was transformed to the rest frame so that the Mg II line at 2798 Å could be accurately located. Two wavelength regions free of Mg~II emission (approximately 2300--2750\,\AA\ and 2850--3300\,\AA\ in the rest frame) were then used to fit a combined baseline model consisting of a power law (representing the jet continuum), a broadened Fe~II pseudo-continuum template \citep{vestergaard2001empirical}, and a constant offset. The \citet{vestergaard2001empirical} UV Fe~II template spans the rest-frame range 2200--3090\,\AA, and in our implementation it is broadened by convolution with a Gaussian kernel corresponding to the instrumental resolution (FWHM $\simeq 1.4$\,\AA\ for SDSS with $R \sim 2000$ and $\simeq 5.6$\,\AA\ for Seimei/VPH-blue with $R \sim 500$). This fit was performed by weighted least squares, using the pixel‐by‐pixel flux uncertainties as weights. After obtaining the best-fit baseline over the full wavelength range, it was subtracted from the observed flux to isolate the Mg II emission residual. The broad emission line profile of Mg II is significantly contaminated by residuals from the strong airglow emission line O I $\lambda$5577. To eliminate the influence of these residuals on the fitting, we masked the spectral region $\lambda_{obs}$ = 5566–5588Å (corresponding to $\lambda_{rest}$ = 2787–2798 Å). Within the 2750–2850 Å interval, the residual spectrum was fitted with two concentric Gaussian components to capture both the narrow and broad wings of the Mg II line. The total Mg II line flux was computed by analytically integrating the two Gaussian profiles, and the rest‐frame EW was obtained by dividing this integrated line flux by the baseline model’s continuum level at 2798 Å. However, the continuum baseline can also be significantly affected by the observing conditions of the standard star used for flux calibration. In particular, if the standard star was observed at a different time or airmass than the science target (due to observation arrangements), the derived response function may introduce residual curvature in the calibrated spectrum. Such mismatches are the most likely cause of the small baseline deviations occasionally seen on the red side of Mg\,II (like T+5.5 and T+8 in Figure \ref{fig:EW}). Moreover, for the flare-phase spectra obtained with KOOLS-IFU, the Mg II emission line is nearly buried in the continuum, causing direct fits to have excessively large uncertainties. To derive physically meaningful measurements—and given that the flux increase during this flare is more likely due to the jet’s synchrotron continuum rather than the BLR—we assume that the BLR-origin Mg II line profile does not change with jet activity. Therefore, in the double-Gaussian fits to the flare spectra, the line centers and widths are fixed to the SDSS values ($\mu$ = 2798 Å; $\sigma$ = 5.79 Å and 20.7 Å), and only the amplitudes are allowed to vary. To estimate the 1$\sigma$ uncertainty on EW, we generated 500 Monte Carlo realizations of the original spectrum by adding Gaussian noise to each pixel according to its measured uncertainty. For each realization, the same baseline‐subtraction and double‐Gaussian fitting procedure was repeated, yielding a distribution of 500 EW values. The standard deviation of that distribution was taken as the 1$\sigma$ error. This Monte Carlo approach naturally accounts for covariances between the continuum‐fit parameters and the Gaussian‐fit parameters, providing a robust estimate of the measurement uncertainty.

As shown in Figure \ref{fig:EW}, the Mg II EW in the SDSS spectrum of OP 313 during its quiet period is significantly larger than the EWs measured during the current flare. In particular, in the first post-flare spectrum (T + 2.5), the Mg II EW reaches its minimum value of only 0.16 ± 0.13 Å. This indicates that the luminosity increase in this flare arises predominantly from the continuum rather than from the emission line, causing the Mg II feature to be effectively “drowned out.” In other words, during the flare OP 313 temporarily enters a ``BL Lac–like" state. The EWs and their uncertainties for all spectra are listed in Table \ref{tab:sed} for reference.
\clearpage

\begin{figure*}[htbp]
\gridline{%
    \fig{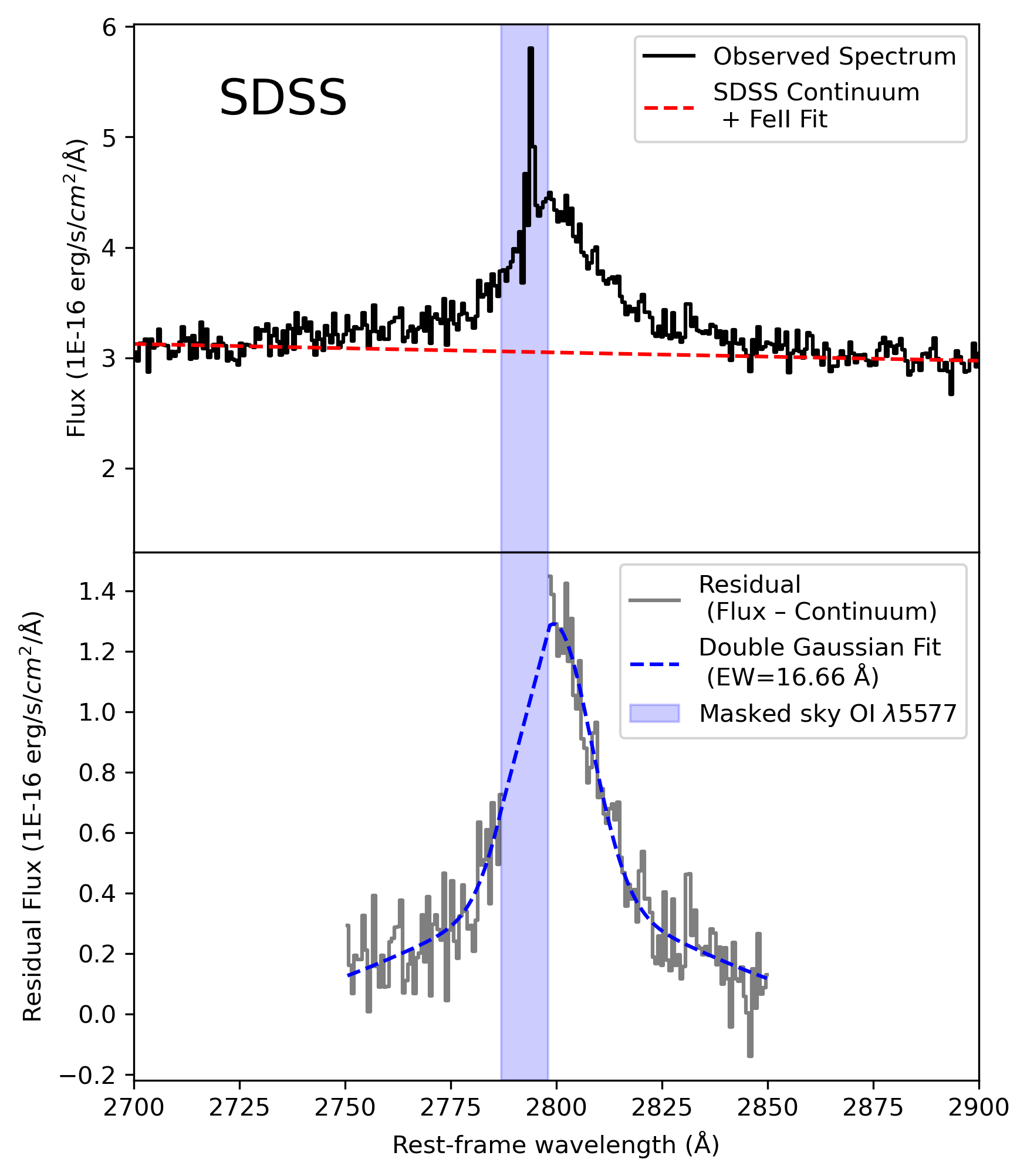}{0.49\textwidth}{}%
   \fig{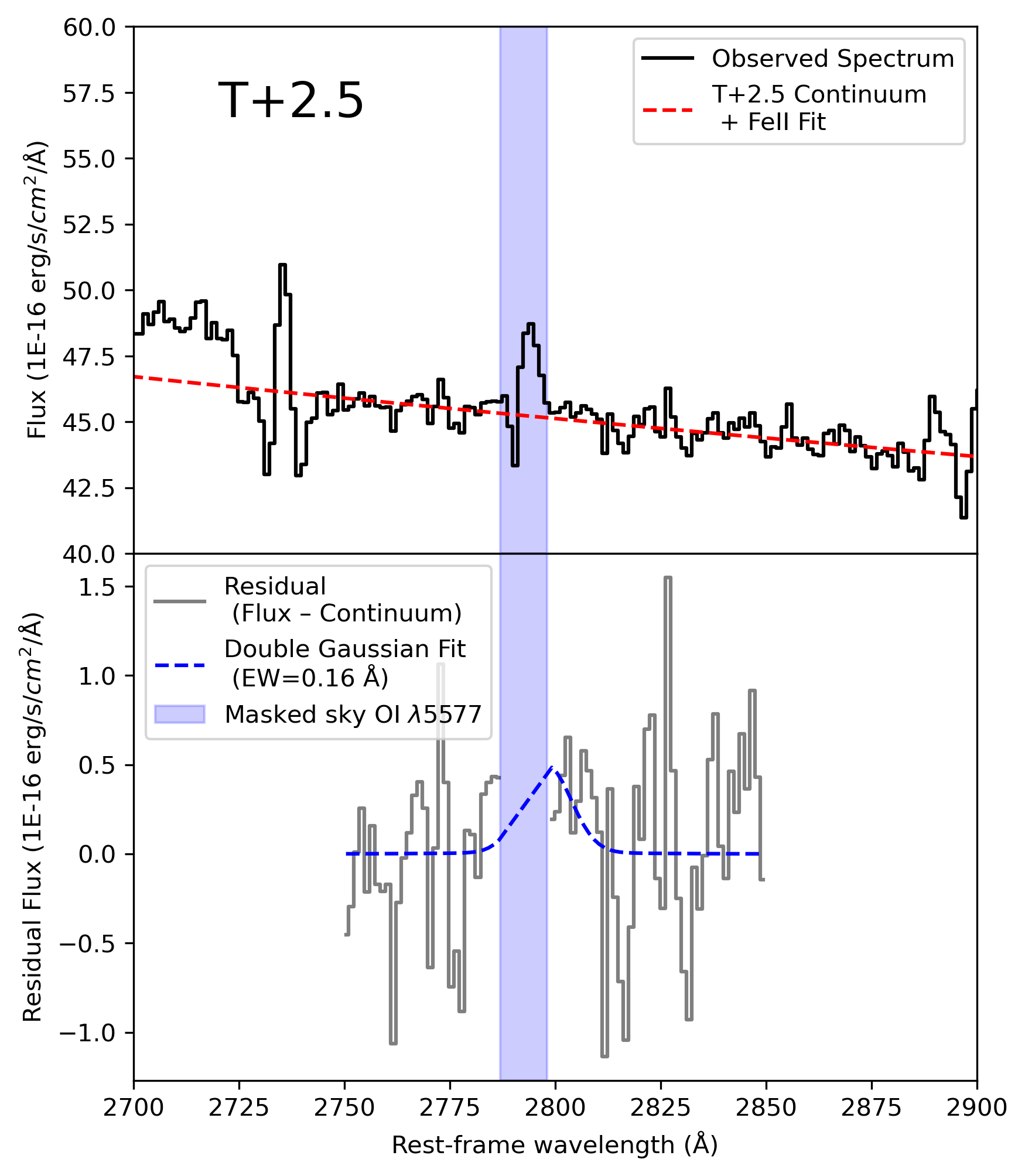}{0.49\textwidth}{}%
   }
\gridline{%
    \fig{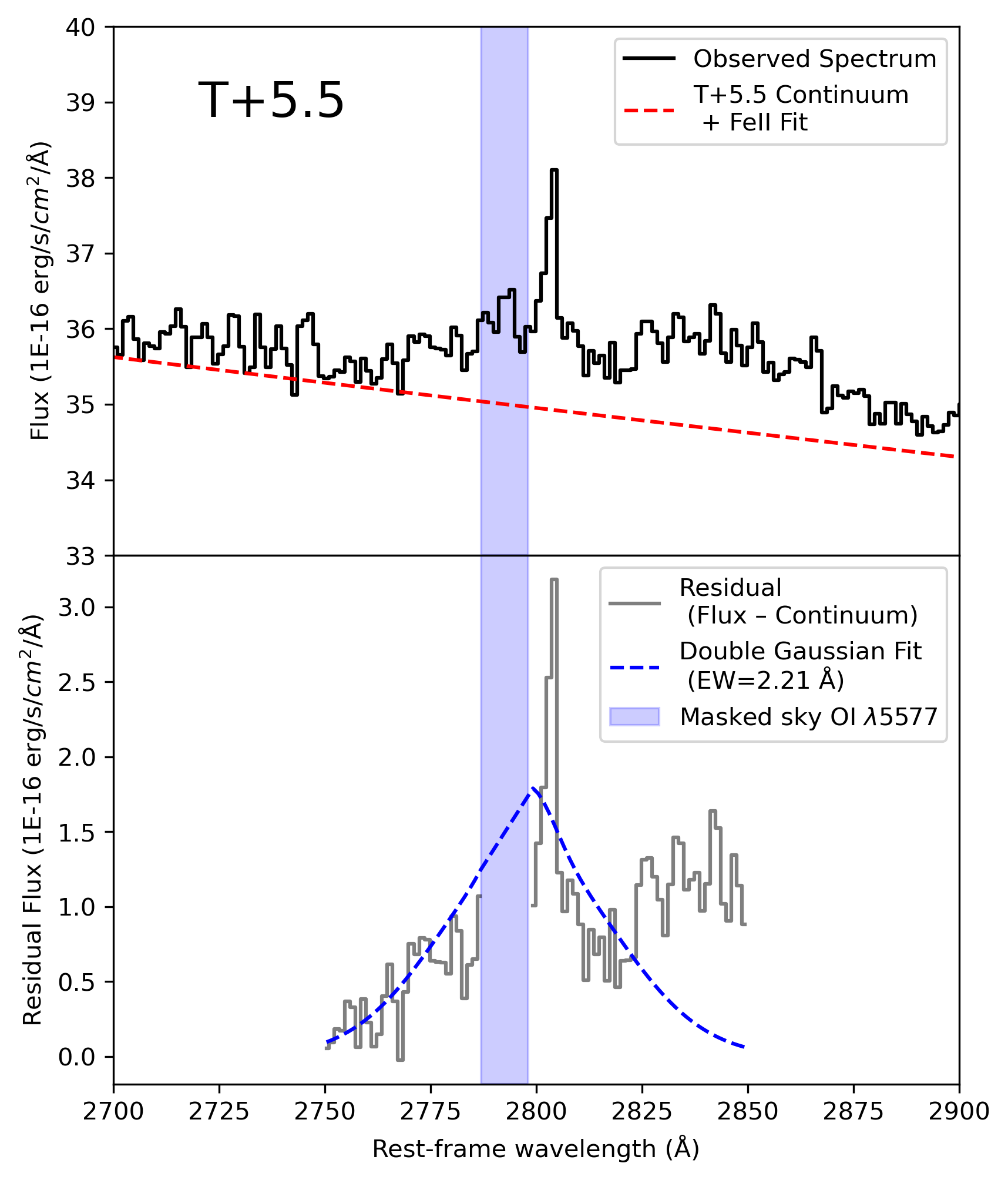}{0.49\textwidth}{}%
   \fig{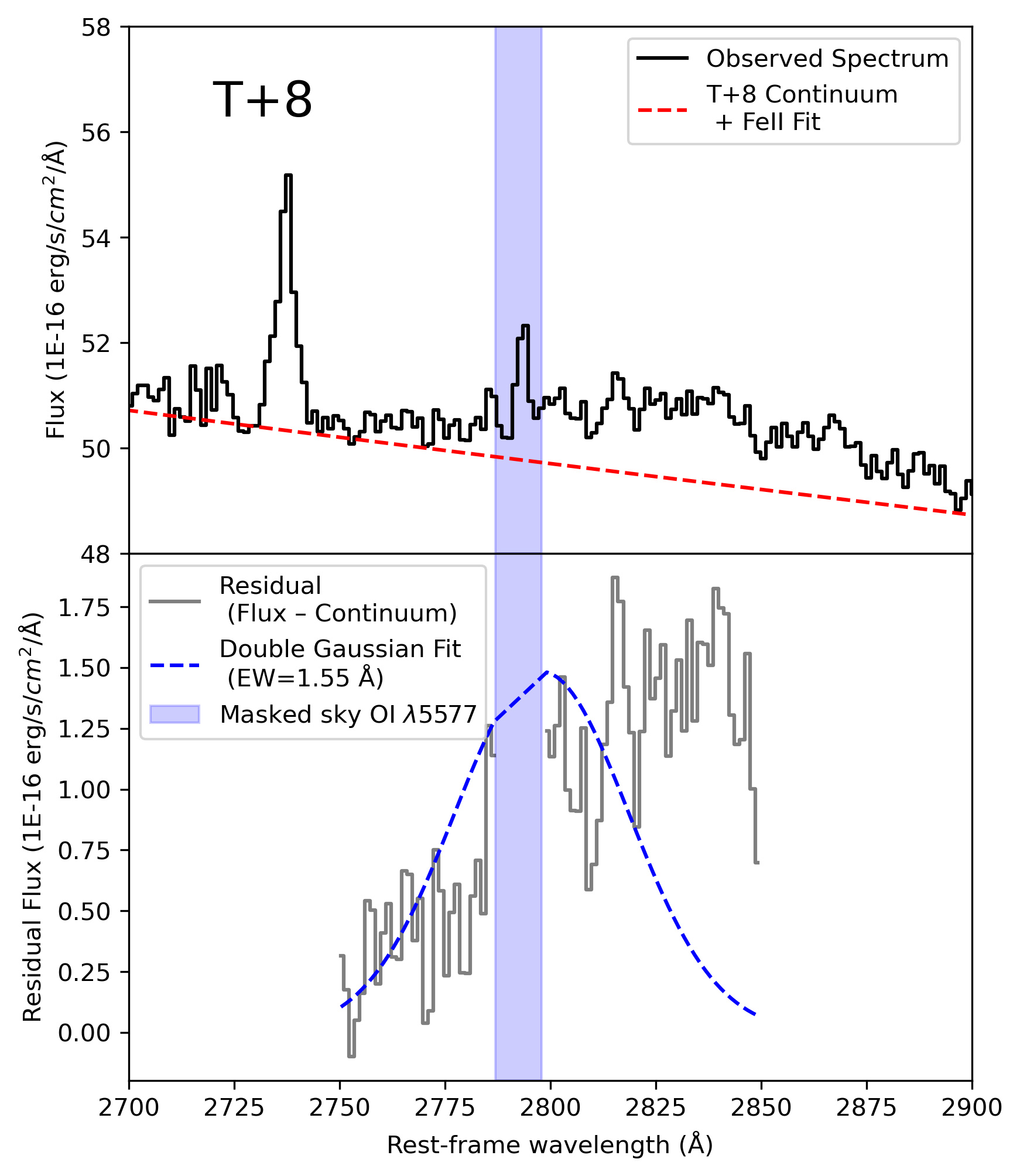}{0.49\textwidth}{}%
}
\end{figure*}
\clearpage
\begin{figure*}[htbp]
\gridline{%
    \fig{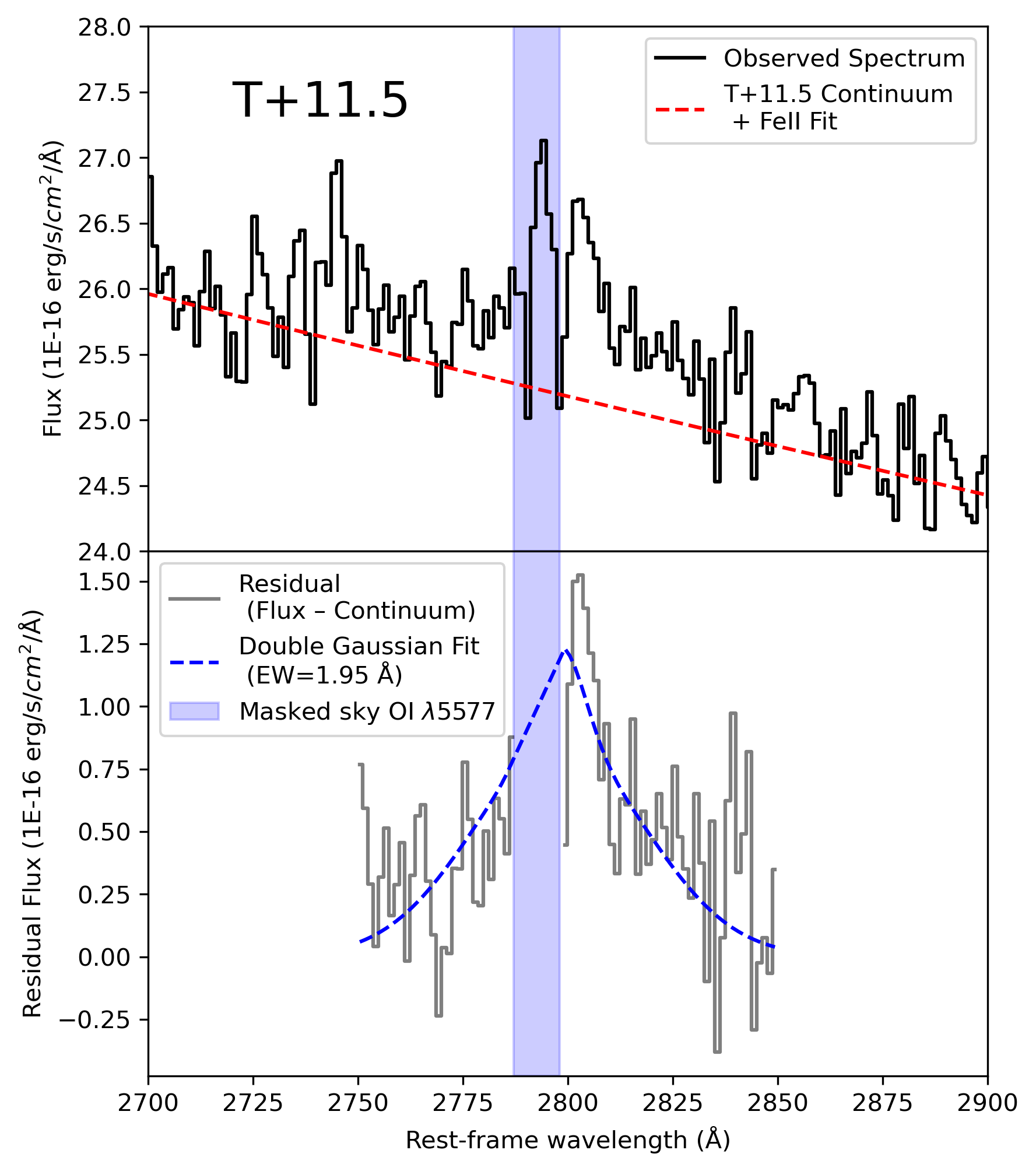}{0.49\textwidth}{}%
   \fig{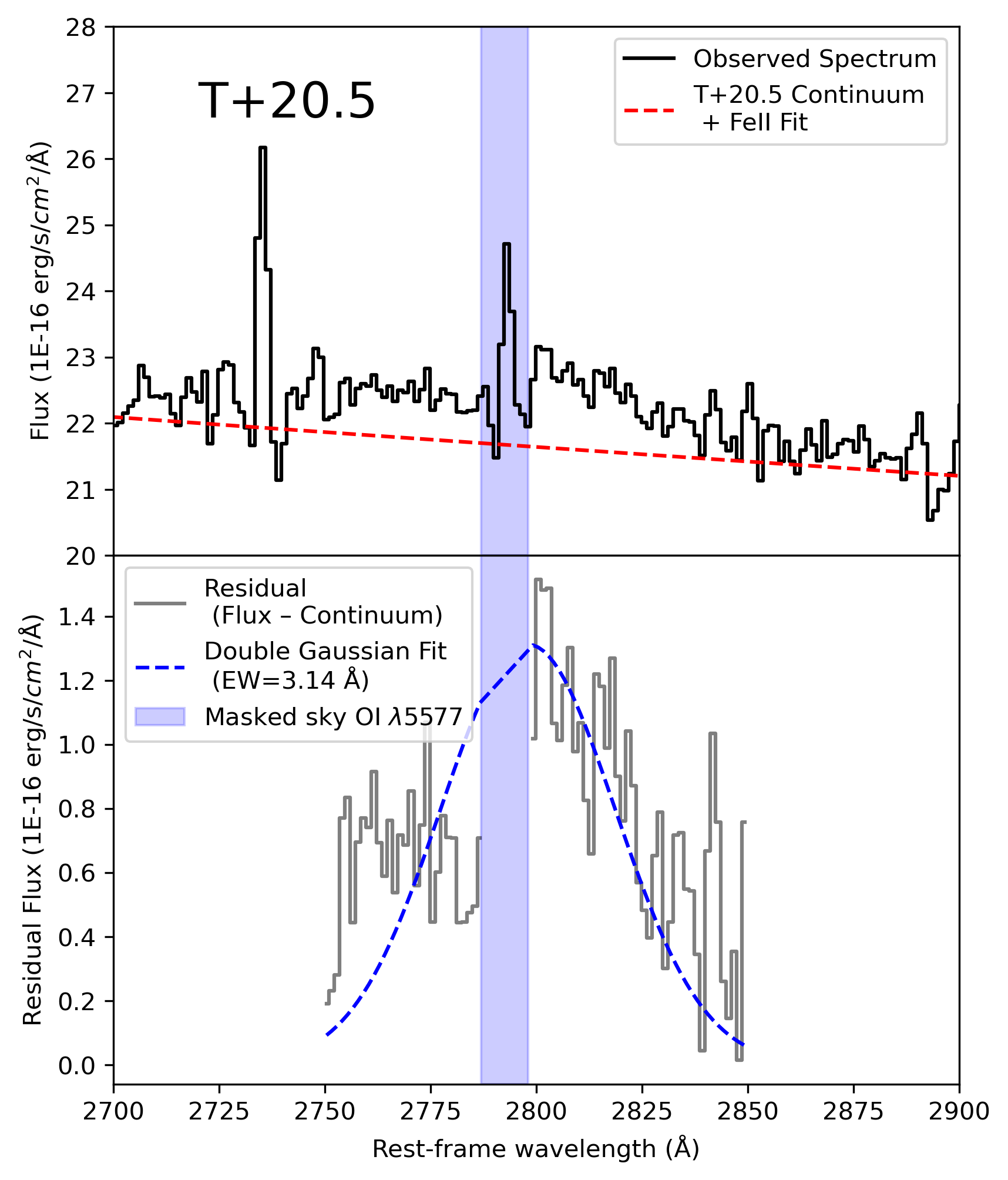}{0.49\textwidth}{}%
   }
\gridline{%
    \fig{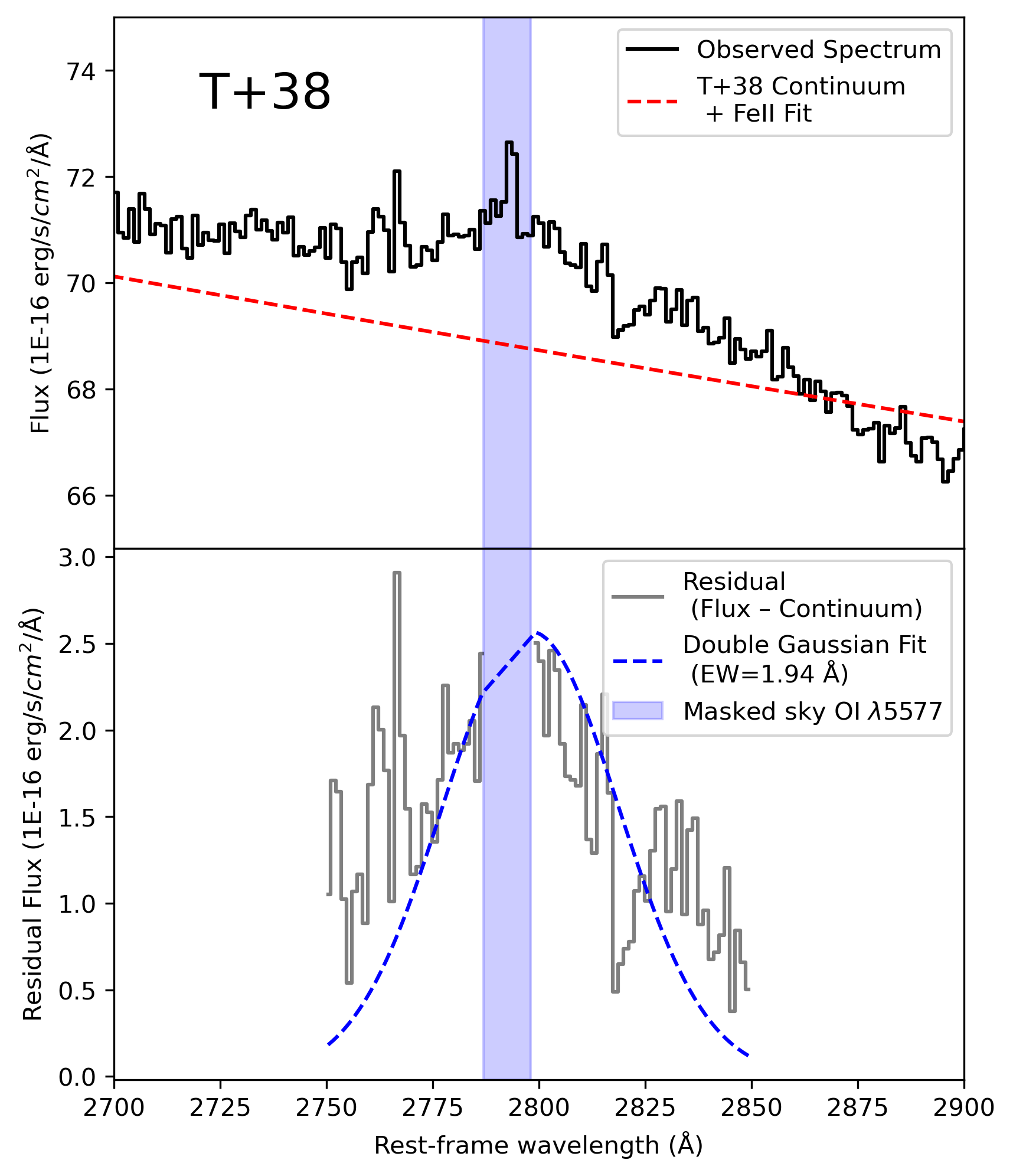}{0.49\textwidth}{}%
   \fig{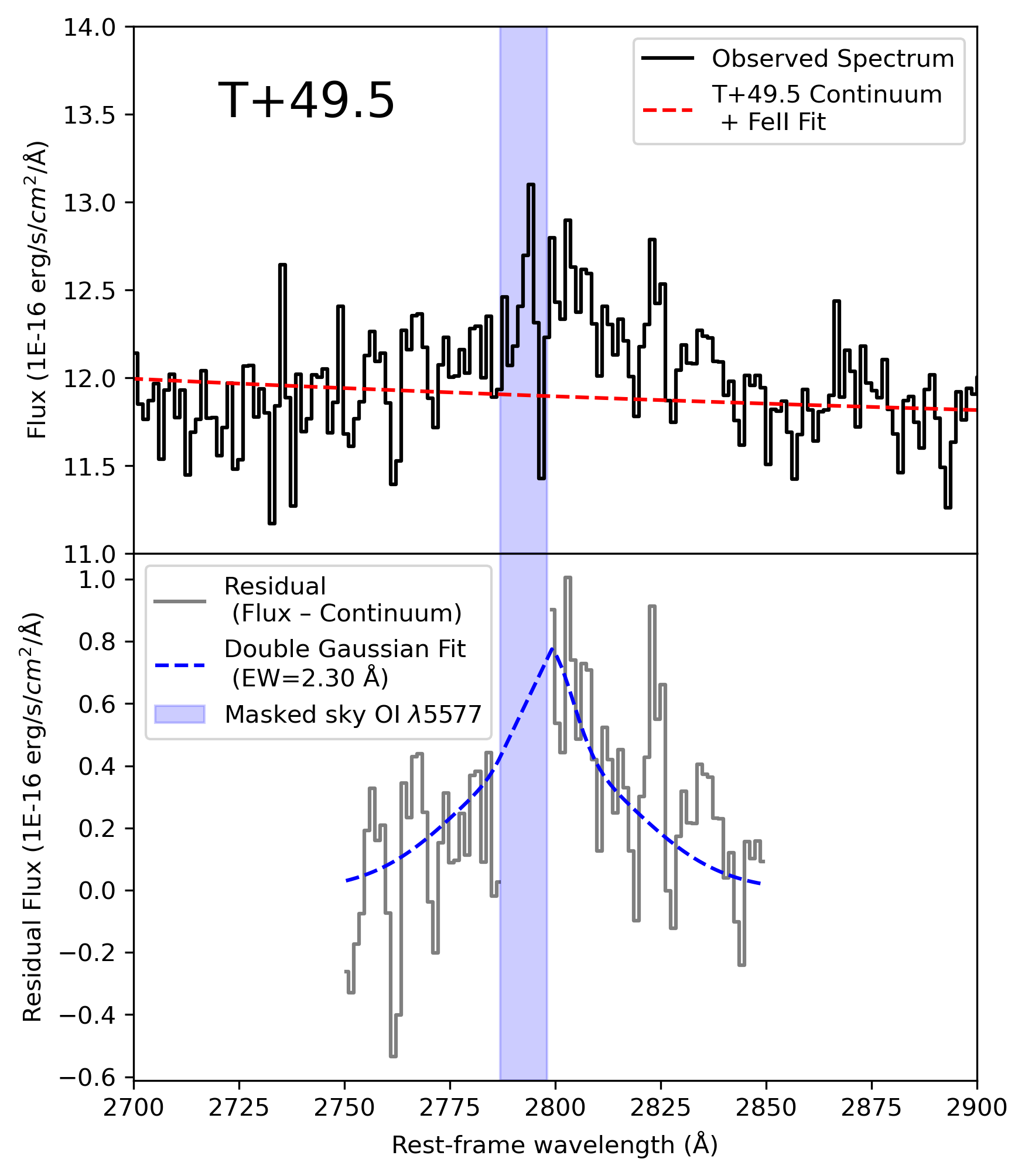}{0.49\textwidth}{}%
     }
  \caption{The fitting results of the Mg II emission line in the OP313 spectra, after removing contributions from the continuum and Fe II emission.}
  \label{fig:EW}
\end{figure*}

\clearpage
On the other hand, from the perspective of the spectral energy distribution, the spectrum obtained at T+2.5, immediately after the $\gamma$-ray flare, is noticeably bluer than all later spectra. This hardening implies that the optical photon index, $\Gamma_O$, decreased during the flare, matching the contemporaneous drop in the $\gamma$-ray photon index, $\Gamma_\gamma$. The coordinated change suggests that the synchrotron peak shifted to higher energies at the flare epoch. Starting at T+5.5, $\Gamma_O$ gradually returned toward its pre-flare value, indicating that the hardening persisted for roughly 5–11 day in the observer frame ($\approx$ 2.5–5.5 d in the rest frame). In terms of flux, the T+5.5 spectrum is about 20$\%$ fainter than that at T+2.5. Even so, two later epochs (T+8 and T+38) show strong optical brightenings that exceed the T+2.5 flux, yet neither event is accompanied by a significant change in $\Gamma_O$. Consequently, our monitoring reveals two clear peaks in optical luminosity: the first at T+2.5, coincident with the $\gamma$-ray flare, and the second at T+38. The expected brightening at T+8 was not confirmed by MITSuME observations because of unfavorable weather conditions.

\begin{figure}[ht!]
    \centering
    \includegraphics[width=0.95\textwidth]{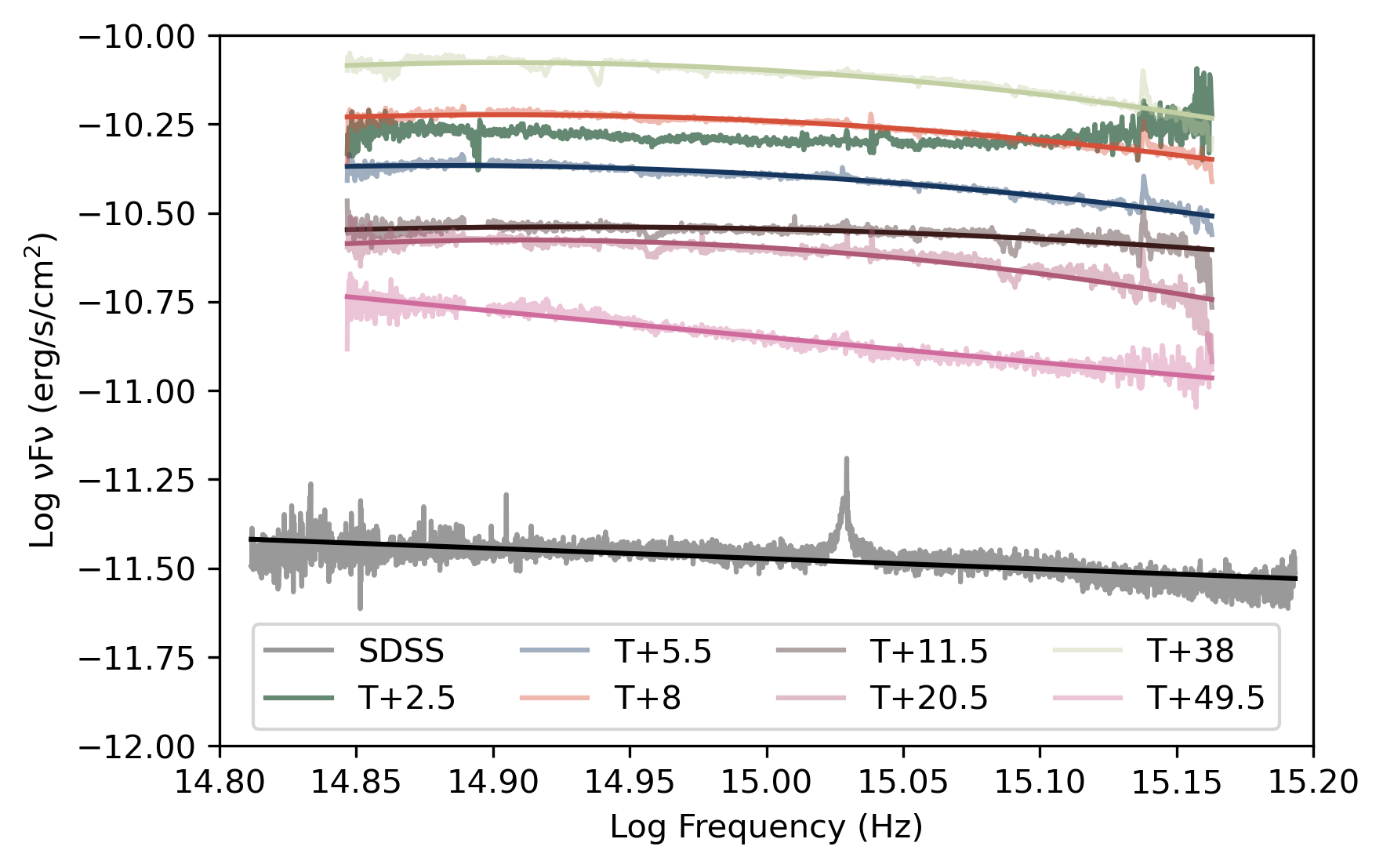}
    \caption{Spectral energy distribution of OP313 after the flare event. Except for the SED fitting at T+2.5, which exhibits a distinct trend, all other SEDs are fitted with a second-order polynomial, represented by solid lines. Non-physical absorption lines affecting the fits have been corrected in this process.}
    \label{fig:op313sed}
\end{figure}

To further investigate the behavior of the synchrotron emission from OP313 after the flare event, we converted the observed spectra into SEDs. Figure \ref{fig:op313sed} shows these SEDs, fitted with a second-order polynomial function derived from the KOOLS-IFU spectra.

In Figure \ref{fig:op313sed} we corrected only the telluric $O_2$ A-band centred at 760 nm, which spans 755–770 nm in our spectra and is unrelated to the blazar. For every epoch we proceeded identically: (i) two clean continuum windows, 735–750 nm and 775–790 nm, were selected; (ii) a linear function, $F_\lambda = a + b\lambda$, was fitted to the fluxes in these windows by ordinary least squares, yielding the coefficients a and b; (iii) the observed fluxes within 755–770 nm were replaced by the values of this fitted line, while the wavelength grid itself was left unchanged. Outside this interval the spectrum was used exactly as recorded. The corrected spectrum was then converted to $\nu F_\nu$ and fitted. No other absorption features were masked or modified. Although minor calibration uncertainties of the standard star could affect individual narrow lines, they do not alter the overall spectral curvature, so our estimates of the synchrotron peak frequency $\nu_s$ and its peak flux $F_s$ remain robust.

Because the signal-to-noise ratio drops sharply at long wavelengths, systematic errors dominate and lead to an unphysical upturn in the fit. The fit therefore yields a positive quadratic coefficient for T+2.5, making the parabola open upward and eliminating any peak within the observed band. A synchrotron SED produced by a single electron population must show downward curvature in the $\log\nu–\log(\nu F_\nu)$ plane, so a positive quadratic coefficient is unphysical. Thus we do not use the formal fit in our analysis. Instead, we treat the T + 2.5 SED qualitatively, note its overall bluer slope, and conclude that the synchrotron peak frequency $\nu_s$ is higher than our observational limit of $1.46 \times 10^{15}\,\mathrm{Hz}$.

Between T+2.5 and T+38, the SEDs display a turning point within our observational range, suggesting that $\nu_s$ may lies within this range during that period. By T+49.5, as well as during the earlier “normal state” represented by the SDSS spectrum, the measured $\nu_s$ falls below our lower observational limit (7.03 $\times$ 10$^{14}$ Hz). These findings illustrate the evolution of $\nu_s$ from higher values observed  following the flare, back down to the lower values characteristic of the silent period.

Table \ref{tab:sed} compiles the synchrotron peak frequencies and fluxes for T+5.5$\sim$T+38, together with their 1 $\sigma$ uncertainties, which were obtained by fitting a log-parabolic model in a mean-centred log-frequency frame, imposing physically motivated bounds on the curvature parameter, and propagating the full coefficient covariance matrix (verified with Monte-Carlo resampling) to each derived quantity; the table thus offers a quantitative view of how these parameters evolved after the flare.

\begin{table}[ht!]
  \caption{The synchrotron radiation parameters of OP313 in different time periods}
  \begin{center}
    \begin{tabular}{c|cccc}
    \hline
    Source & $\nu_s$ (Hz) & Log $\nu_s$ (Hz) & $F_s$ (erg/s/cm$^2$)& $EW_{Mg II}(\AA)$ \\
    \hline
    4LAC & 5.31e+12 & 12.7250 &  5.82e-12 & --\\
    SDSS & $<$6.48e+14 & $<$14.8116 & $>$0.32e-11 & 16.66 ± 3.77\\
    \textcircled{1}T+2.5&  $>$14.60e+14 & $>$15.1644 & $>$6.31e-11 & 0.16 ± 0.13\\
    \textcircled{2}T+5.5& 7.62e+14 ± 3.60e+12 &14.8822 ± 0.0020 & 4.30e-11 ± 3.69e-14 &  2.21 ± 0.08\\
    \textcircled{3}T+8& 8.02e+14 ± 2.98e+12 & 14.9044 ± 0.0016 & 5.98e-11 ± 4.11e-14 & 1.55 ± 0.07\\
    \textcircled{4}T+11.5 & 8.50e+14 ± 6.45e+12 & 14.9293 ± 0.0033 & 2.89e-11 ± 3.07e-14 & 1.95 ± 0.14\\
    \textcircled{5}T+20.5 & 8.12e+14 ± 4.16e+12 & 14.9095 ± 0.0022 & 2.66e-11 ± 3.55e-14 &3.14 ± 0.16\\
    \textcircled{6}T+38 & 8.04e+14 ± 3.35e+12 & 14.9052 ± 0.0018 & 8.38e-11 ± 8.11e-14 & 1.94 ± 0.07\\
    \textcircled{7}T+49.5 & $<$7.03e+14 & $<$14.8470 & $>$1.84e-11 & 2.30 ± 0.21\\
    \hline
    \end{tabular}%
      \end{center}
  \label{tab:sed}%
\end{table}%

Our measurements show that the synchrotron peak frequency of OP313 rose from $\approx 6 \times 10^{12} Hz$ to  $\approx 8 \times 10^{14} Hz$ after the flare about two orders of magnitude higher than the single value tabulated for this source in the 4LAC catalogue.  Such a large shift is not unexpected for changing-look blazars: the time-dependent jet model of \cite{pandey2024origin} predicts that $\nu_s$ can move from $\approx 8 \times 10^{12} Hz$ to $\approx 6 \times 10^{14} Hz$ as the electron population hardens during a major outburst.  The range we observe is therefore fully consistent with theoretical expectations and confirms that the 4LAC value represents a long-term quiescent state rather than the flaring conditions captured in our campaign.

Unlike the result of \cite{pandey2024origin}, we detected a much higher synchrotron peak immediately after the flare (T + 2.5): $\nu_s > 1.46 \times 10^{15} Hz$ . \cite{pandey2024origin} used broadband data that are typically separated by one or several weeks, so their fitted value of $\nu_s \simeq 6 \times 10^{14}Hz$  represents a time-averaged peak for a long-duration outburst and cannot track the rapid, day-scale variations that follow the flare maximum. Our measurement at T + 2.5 therefore captures the short-lived “upper envelope” produced during an extreme particle-acceleration episode, where $\nu_s$ rises above $1.46 \times 10^{15} Hz$. The gap between the two peak frequencies can be fully explained by a modest increase in either the characteristic electron Lorentz factor or the Doppler boosting factor during the very brightest phase, so the two results are not contradictory. These observations imply that OP313 can momentarily enter the high synchrotron peaked regime at the flare maximum and then quickly relax to the intermediate synchrotron peaked state described by \cite{pandey2024origin}.
\clearpage
\section{Discussion}
\label{sec:Discussion}
The above observations suggest that OP313 is a transitional blazar, similar to the example presented by \cite{ruan2014nature}. By examining its spectra and multi-wavelength light curves (Figure \ref{fig:op313sep}), we outline a scenario in which OP313 evolves from an FSRQ-like state into a BL Lac-like state, and then returns to its original state. To describe this process, we divide the event into several chronological phases relative to the time of the flare, denoted as T, with T+X indicating X days after the flare in the rest frame:

Before T-50: Both the $\gamma$-ray energy flux and photon index remained at quiescent levels consistent with the values reported in the 4LAC catalog. We define this quiescent configuration as the ``normal state" of OP313. During this period, no optical data were available due to the target’s visibility constraints. In this state, both inverse Compton (IC) scattering and synchrotron emission were presumably stable, showing no signs of the activity.

T-50 $\sim$ T-1: The $\gamma$-ray energy flux increased by approximately an order of magnitude over a single day, while the $\gamma$-ray photon index dropped rapidly to around 1.7. In the standard one-zone leptonic model, in which the same population of relativistic electrons produces both synchrotron and inverse Compton (IC) emission, an increase in the characteristic electron energy drives both the IC peak frequency ($\nu_{IC}$) and the synchrotron peak frequency ($\nu_{s}$) to higher values \citep{fossati1998unifying}. Observationally, this can manifest as a rise in the IC power ($P_{IC}$) and a decrease in the gamma-ray photon index \citep{abdo2010spectral}. As shown in Figure \ref{fig:op313sep}, at the start of this interval (T–50) the $P_{IC}$ rose by an order of magnitude relative to the normal state, accompanied by a significant decrease in the gamma-ray photon index. Optical observations from Tomo-e Gozen at T-40 recorded a magnitude of about 14, comparable to the value at T=0, with some fluctuations over the following 40 days. These data indicate that after T–50 both inverse Compton scattering and synchrotron emission entered an active state, with their powers and peak frequencies rising sharply and then fluctuating strongly over the following fifty days. We refer to this phase as the ``pre-transitional phase".

T-1 $\sim$ T+2.5: At T-1, $\nu_{IC}$ increased rapidly, reaching its highest value of the entire transitional event. In contrast, $P_{IC}$ initially remained unchanged, showing a delay similar to what was seen in the ``pre-transitional phase". Only at T=0 did $P_{IC}$ surge dramatically—by a factor of about 50—achieving its peak for the event. After T=0, $P_{IC}$ swiftly declined, followed by a slight reduction in $\nu_{IC}$. By T+2.5, both had returned to levels resembling the ``pre-transitional phase". Throughout this interval, synchrotron power ($P_{s}$) and $P_{IC}$ were positively correlated. Both peaked at T=0 and then decreased, with minor fluctuations, until about T+15. Meanwhile, $\nu_{s}$ remained significantly elevated until T+2.5, subsequently returning to near-normal values within the next few days. We refer to this phase as the ``transitional state."

T+2.5 $\sim$ T+49.5: $\nu_{IC}$ and $P_{IC}$ reverted to a relationship similar to that seen in the pre-transitional state, again showing a positive correlation. The average value of $\gamma$-ray photon index settled around 2.1, an intermediate level between the normal and pre-transitional states, and it did not drop as low as the 1.8 value recorded earlier. The peak $P_{IC}$ during this ``post-transitional phase" was also intermediate, lying between the ``normal" and ``pre-transitional" values. This indicates that while IC activity persisted, it was diminishing overall. The positive correlation between $P_{s}$ and $P_{IC}$ continued until T+15. Afterward, $P_{s}$ underwent additional fluctuations and eventually rose again, nearing its earlier peak at around T+40, suggesting a distinct episode of synchrotron activity not necessarily directly tied to the flare-driven events. Throughout this post-transitional state, $\nu_{s}$ remained relatively stable, showing no major shifts.

After T+49.5: The optical spectrum of OP313 had finished to revert from its BL Lac-like appearance back toward an FSRQ-like spectrum, completing a full transition cycle. 

The Fermi/LAT energy flux is positively correlated to the inverse Compton scattering power ($P_{IC}$) of OP313 in the $\gamma$-ray band, whereas the $\gamma$-ray photon index is negatively correlated with the peak frequency of the inverse Compton component ($\nu_{IC}$) in the spectral energy distribution (SED) of OP313\citep{abdo2010spectral}. Approximately 100 days before the flare, OP313 entered a relatively active $\gamma$-ray state that persisted for more than 100 days following the flare. We infer that this flare and subsequent transitions between FSRQ-like and BL Lac-like states are driven by variations in the accretion rate of the accretion disk. Changes in the accretion rate affect the efficiency with which electrons migrate from the black hole’s equatorial regions to the polar zones and into the jet, propelled by the black hole’s magnetic fields \citep{wald1974black,tchekhovskoy2011efficient}. This process ultimately triggered the described transition event.

Moreover, although in certain special cases the $\gamma$-ray activity may arise from enhanced  External Compton (EC) in the accretion flow rather than the jet \citep{dermer1993model, sikora1994high}, purely EC-driven $\gamma$-ray flares necessarily entail an overall decrease in the synchrotron power due to stronger IC cooling \citep{ghisellini2009canonical}. This is inconsistent with our optical observations before and after the flare (both Tomo-e Gozen and HONIR). Therefore, we conclude that the current $\gamma$-ray flare is dominated not by EC but by Synchrotron Self-Compton (SSC) in jet.

As the number of electrons injected into the jet by magnetically driven processes rises, OP313 enters the ``pre-transitional state". The increase in the electron population participating in both synchrotron radiation and IC scattering boosts $P_s$ and $P_{IC}$. After approximately 50 days in this ``pre-transitional state", $P_{IC}$ reaches its highest value of the entire event, while $P_s$ remains near its previously established peak. We suspect that the limitations in real-time monitoring by Tomo-e Gozen and the delayed start of MITSuME observations after the Fermi/LAT ToO alert may have limited our ability to capture the exact temporal evolution of $P_s$. 

Furthermore, it is worth noting that \cite{pandey2024b2} also obtained spectroscopic observations of OP313 at T+56 using the CAFOS instrument on the 2.2 m telescope at the Centro Astronómico Hispano en Andalucía (CAHA). They detected a clear Mg II emission line in that spectrum and derived an Eddington ratio of 0.62 ± 0.43 based on its line profile, which exceeds the commonly adopted ``a few$\times 10^{-3}$'' threshold \citep{ghisellini2008blazar} distinguishing BL Lac and FSRQ states. This result is consistent with our suggestion that OP313 continued to evolve toward an FSRQ-like state after T+50. However, during the transition when OP313 exhibited a BL Lac–like state, the Mg II line was effectively drowned out by the enhanced synchrotron continuum, making it extremely difficult to measure a reliable line profile and thus calculate a valid Eddington ratio. Consequently, the approach used by \cite{pandey2024b2} cannot be applied to an object actively undergoing this state transition.

We have thus provided a detailed account of an event in which OP313 transitions from an FSRQ-like state to a BL Lac-like state, and subsequently returns to an FSRQ-like state. By analogy, the reverse transitions described in earlier studies, where BL Lacs develop FSRQ-like characteristics, may also be driven by variations in the accretion rate. A lower accretion rate would reduce the number of high-energy electrons injected into the jet, thereby decreasing $P_s$. In this scenario, the reduced electron density diminishes the synchrotron emission and more readily reveals the broad-line region emission lines, as illustrated in \cite{ruan2014nature} (their Figure 5).

\begin{figure}[ht!]
    \centering
    \includegraphics[width=\textwidth]{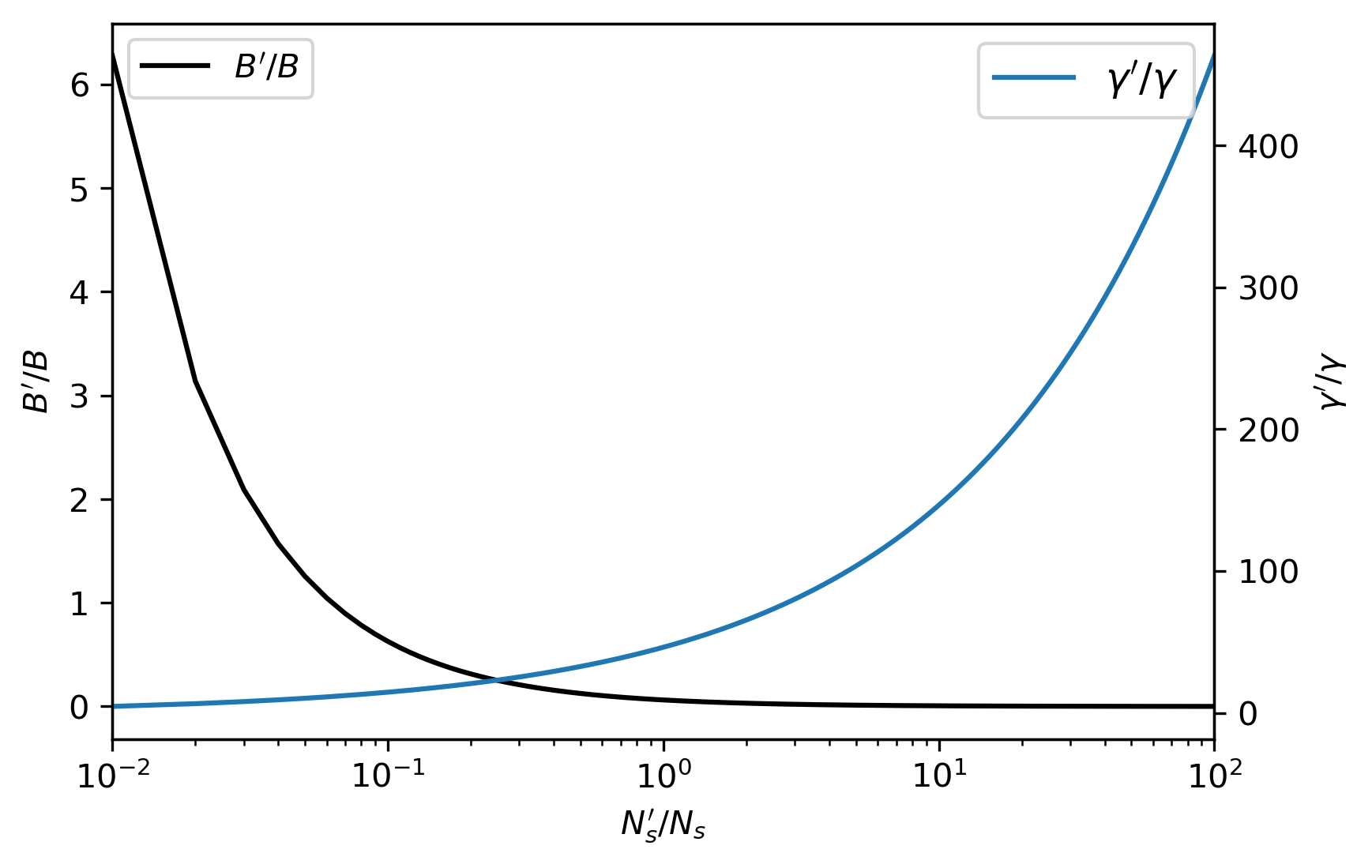}
    \caption{Order estimate of $\gamma'$ and $B'$ of OP313 between T+5.5 and T+38. In the figure, the rate of change of the magnetic field is represented by the left axis, while the rate of change of the particle Lorentz factor is represented by the right axis. The primed quantities correspond to values in the ``post-transition state" (T + 5.5 to T + 38), while the unprimed quantities represent the ``normal state", as listed in the 4LAC catalog.}
    \label{fig:ns}
\end{figure}

To further quantify the conditions within the shock front (i.e. the narrow, high-pressure zone that forms where the propagating disturbance strongly compresses the jet plasma) during the ``post-transitional state", we refer to the data in Table \ref{tab:sed}. To determine the synchrotron peak parameters, we first converted the observed flux density per unit wavelength, $F_\lambda$, into flux density per unit frequency using 
$F_\nu = F_\lambda \lambda^2 / c$, with $\nu = c/\lambda$. We then constructed the SED in the form $\nu F_\nu$, which represents the emitted power per logarithmic frequency interval. For each spectrum, we fitted the $\log \nu$--$\log (\nu F_\nu)$ relation with a quadratic function and identified the vertex of the parabola as the synchrotron peak. The frequency at the vertex corresponds to the synchrotron peak frequency $\nu_s$, and the ordinate gives the peak power $p_s \equiv (\nu F_\nu)_{\rm peak}$. In the one-zone synchrotron self-Compton model (without EC), the synchrotron peak frequency $ \nu_s $ scales with the electron Lorentz factor $ \gamma $, magnetic field strength $ B $, and bulk Doppler factor $ \delta $ as $ \nu_s \propto \gamma^2 B \delta $, while the peak power $ P_s $ scales with the electron density $ N_s $, $ \gamma $, $ B $, and $ \delta $ as $ P_s \propto N_s \gamma^2 B^2 \delta^4 $. By comparing the ratios $ \nu_s’/\nu_s $ and $ P_s’/P_s $ between the flaring (T+5.5 to T+38) and quiescent states (e.g., the 4LAC and SDSS periods), and assuming that the bulk Doppler factor remains unchanged in the absence of strong geometric effects, one can derive the magnetic field and electron energy for different electron densities $N_s$. For example, $ B’/B = (P_s’/P_s)  (\nu_s’/\nu_s)^{-1} (N_s’/N_s)^{-1} $ and $ \gamma’/\gamma = (\nu_s’/\nu_s) \sqrt{(N_s’/N_s) / (P_s’/P_s)} $, thereby quantitatively characterizing the impact of the flare event on the jet’s physical parameters. Also, to illustrate the coupled evolution of the magnetic field and characteristic electron energy under different particle density variations, we sample $N_s’/N_s$ from 0.01 to 100 in steps of 0.01 which covers the range predicted by most past studies \citep{krawczynski2002time}. Smaller ratios ($N_s’/N_s \lesssim 0.1$) correspond to scenarios dominated by particle escape or cooling, while larger ratios ($N_s’/N_s \gtrsim 10$) represent strong particle injection or acceleration events (e.g. OP313 flare event). These estimates, in conjunction with various assumed ratios of $N_s'/N_s$, are depicted in Figure \ref{fig:ns}.

\clearpage

From Figure \ref{fig:ns}, it is clear that unless the electron density in the emission zone between T+5.5 and T+38 is reduced to less than six percent of its ``normal-state" value, the magnetic field strength during ``post-transitional state" must be lower than that in the ``normal state". Such a substantial decrease in electron density conflicts with our earlier inference that an increase in the accretion rate triggers the transitional event. Therefore, we infer that the magnetic field strength within the shock front declines during the ``post-transitional state." Such behavior has also been predicted by earlier models. In the \cite{marscher1985models} framework, the jet undergoes an adiabatic expansion phase during which the plasma in the shock‐compressed region rapidly increases in volume. Under the frozen‐in condition of the magnetic field, the local field strength then decays as a power law of the characteristic size (approximately $B\propto R^{-1\sim2}$), naturally reducing the initially high $B$ value. \cite{mimica2009spectral}  found that while the magnetic field is compressed and amplified during the initial shock formation, as the shock propagates into the downstream slower region, turbulence, magnetic reconnection, and volume expansion transfer magnetic energy to the particles and carry it further downstream, causing the field strength to decay along the jet axis approximately as $B\propto x^{-1}$. These models consistently show that in the later stages of an internal shock, plasma expansion and energy dissipation lead to a decline in the effective magnetic field strength, in agreement with our observations.

Another noteworthy phenomenon is the breakdown of the previously observed correlation between synchrotron and IC scattering after the flare. A comparison of the optical and $\gamma$-ray light curves during the flare event (Figure \ref{fig:op313sep}(B)$\&$(E) shows details about this change. As shown in the figure, before the flare, the synchrotron and IC emission appeared to track each other closely. However, during the event, OP313 exhibited three optical peaks at MJD 60290, 60369, and 60445. The first two peaks coincided with $\gamma$-ray peaks, suggesting that both synchrotron and IC components were still evolving in concert during the early portion of the flare. However, there was no corresponding $\gamma$-ray peak coincident with the highest optical peak at MJD 60445. This indicates that, following the flare, the synchrotron emission could increase independently, without a commensurate rise in IC scattering power. Prior to the flare described in this study, OP313 showed no comparably large events in its Fermi/LAT detection history (2008–present). Likewise, the survey cadence of Tomo-e Gozen (and other programs such as ZTF)—as seen in the portion of Figure \ref{fig:op313sep}(D) before time T—is much lower than that of Fermi/LAT (minimum three days). This discrepancy prevents us from conducting a statistically meaningful retrospective comparison using archival survey data on such short- timescale event.

However, similar cases have been reported in the literature: \cite{rajput2019temporal} identified four major optical flaring epochs of 3C 454.3 over 9 years (2008–2017). In their epoch D (around MJD 66550), the source exhibited large-amplitude optical flares while corresponding $\gamma$-ray flares were weak or absent. \cite{nesci2021multiwavelength} likewise reported an optical brightening near MJD 56890 in S5 1803+784 that appeared independently of any $\gamma$-ray activity. These precedents show that the optical enhancement observed in OP313 is not unique. Comparable orphan optical outbursts have also been documented in PKS~0208$-$512, most notably during MJD~55087--55233 (2009 September to 2010 February), when a bright optical/near-IR flare showed no $\gamma$-ray counterpart \citep{chatterjee2012optical}, and in S4~1849+67, where a broad optical flare around MJD~56100 likewise lacked any associated $\gamma$-ray activity \citep{cohen2014temporal}.

One possible explanation for this phenomenon is the IC component of blazar emission consists of both Synchrotron Self-Compton (SSC) and External Compton (EC) processes. SSC involves synchrotron photons being upscattered by the same high-energy electrons that produced them, making its intensity strongly dependent on the synchrotron emission power. Since the observed decoupling cannot be explained solely by SSC, we must consider EC processes. EC scattering involves seed photons originating from external regions such as the broad line region. If the shock front moves away from these regions or the photon path between the shock front and the broad line region is restricted, the supply of external seed photons diminishes. Under such circumstances, even if SSC increases alongside synchrotron emission, the overall IC power could remain unchanged or decline due to reduced EC contributions.

\section{Conclusions}\label{sec:sum}

Over the course of 100 days following the brightening event on February 27, 2024, we conducted extensive optical spectroscopic and photometric observations of the low $\gamma$-ray photon index FSRQ OP313. Combined with continuous $\gamma$-ray monitoring from Fermi/LAT, these data revealed that OP313 entered a more active state approximately 50 days before the major flare. During this ``pre-transitional state,” both synchrotron emission and inverse Compton scattering intensified. We attribute this change to variations in the accretion rate of the disk, which in turn altered the number and typical energies of electrons injected into the jet, thereby enhancing the radiative output.

After maintaining elevated activity for about 50 days, OP313 was likely to undergo a brief but pronounced flare, with its $\gamma$-ray flux increasing to roughly 100 times the ``normal state” level and about 10 times that of the ``pre-transitional state.” Simultaneously, the synchrotron peak frequency ($\nu_s$) rose by nearly two orders of magnitude. This extreme shift in $\nu_s$ obscured broad-line region features in the optical spectrum, producing a distinctly non-flat continuum that resembled a BL Lac rather than an FSRQ. This intense ``transitional state,” however, lasted less than three days.

Subsequently, OP313 entered a ``post-transitional state” lasting at least 50 days. During this phase, synchrotron emission remained exceptionally strong, with the peak frequency staying over 100 times higher than in the ``normal state” for about 40 days. In contrast, inverse Compton scattering activity waned and became decoupled from the synchrotron component. We interpret this desynchronization as the result of fewer external seed photons available for EC scattering. This reduction in external seed photons created a bottleneck effect, limiting the fraction of electrons participating in IC scattering and contributing to a decline in the power of IC scattering within the shock front. After approximately 50 days in the post-transitional state, the $\nu_s$ decreased significantly, marking the gradual conclusion of the event. However, due to limitations in visibility, we were unable to continue follow-up observations using the Seimei Telescope. Nevertheless, as shown in Figure \ref{fig:op313sed}, the declining $\nu_s$ and $P_s$ suggest that the source is recovering from the effects of the flare-induced transition event and is gradually returning to its original FSRQ-like state.

Our results suggest that some blazars, such as OP 313, can evolve smoothly between the traditional BL Lac and FSRQ classes rather than undergoing discrete jumps. This transition is reflected in several observational parameters, including the $\gamma$-ray photon index, the Eddington ratio, the emission-line equivalent width, and the synchrotron peak frequency. As our understanding of such events improves, it becomes increasingly inappropriate to force transitional sources like OP 313 into either category. More broadly, the conventional BL Lac/FSRQ dichotomy should be replaced by a quantitative classification scheme based on parameters such as the $\gamma$-ray photon index, where non-transitional FSRQs occupy the high-index regime, non-transitional BL Lacs the low-index regime, and transitional blazars trace continuous trajectories between them.

\begin{acknowledgments}
This work was supported by the Japan Society for the Promotion of Science (JSPS) KAKENHI grants (20K14521, 23H04894, 23K03459, 24H00027, 24H01812, 24K00673, 24K07090, 24K07091) and in part by the Optical and Infrared Synergetic Telescopes for Education and Research (OISTER) program funded by the MEXT of Japan. It was also partially carried out by the joint research program of the Institute for Cosmic Ray Research (ICRR), The University of Tokyo. We thank the staff of Okayama Observatory, Kyoto University, and the Okayama Branch Office, NAOJ, NINS, for their help with the KOOLS-IFU observations. We are grateful to Prof. Fumihide Iwamuro from the Department of Astronomy and Astronomy at the Graduate School of Science, Kyoto University, for his assistance with the KOOLS-IFU spectroscopic analysis pipeline. We also thank Prof. Asano Katsuaki of the Institute for Cosmic Ray Research, The University of Tokyo, and Prof. John Silverman of the Kavli Institute for the Physics and Mathematics of the Universe for their help in preparing this article.

\end{acknowledgments}

\bibliography{sample631}{}

\begin{thebibliography}{}
\expandafter\ifx\csname natexlab\endcsname\relax\def\natexlab#1{#1}\fi
\providecommand{\url}[1]{\href{#1}{#1}}
\providecommand{\dodoi}[1]{doi:~\href{http://doi.org/#1}{\nolinkurl{#1}}}
\providecommand{\doeprint}[1]{\href{http://ascl.net/#1}{\nolinkurl{http://ascl.net/#1}}}
\providecommand{\doarXiv}[1]{\href{https://arxiv.org/abs/#1}{\nolinkurl{https://arxiv.org/abs/#1}}}

\bibitem[{Abdo {et~al.}(2010)Abdo, Ackermann, Agudo, Ajello, Aller, Aller, Angelakis, Arkharov, Axelsson, Bach, {et~al.}}]{abdo2010spectral}
Abdo, A., Ackermann, M., Agudo, I., {et~al.} 2010, The Astrophysical Journal, 716, 30

\bibitem[{Ajello {et~al.}(2020)Ajello, Angioni, Axelsson, Ballet, Barbiellini, Bastieri, Gonzalez, Bellazzini, Bissaldi, Bloom, {et~al.}}]{ajello2020fourth}
Ajello, M., Angioni, R., Axelsson, M., {et~al.} 2020, The Astrophysical Journal, 892, 105

\bibitem[{Akitaya {et~al.}(2014)Akitaya, Moritani, Ui, Urano, Ohashi, Kawabata, Nakashima, Sasada, Sakimoto, Harao, {et~al.}}]{akitaya2014honir}
Akitaya, H., Moritani, Y., Ui, T., {et~al.} 2014, in Ground-based and Airborne Instrumentation for Astronomy V, Vol. 9147, SPIE, 1520--1534

\bibitem[{Atwood {et~al.}(2009)Atwood, Abdo, Ackermann, Althouse, Anderson, Axelsson, Baldini, Ballet, Band, Barbiellini, {et~al.}}]{atwood2009large}
Atwood, W., Abdo, A.~A., Ackermann, M., {et~al.} 2009, The Astrophysical Journal, 697, 1071

\bibitem[{Baba {et~al.}(2002)Baba, Yasuda, Ichikawa, Yagi, Iwamoto, Takata, Horaguchi, Taga, Watanabe, Ozawa, {et~al.}}]{baba2002development}
Baba, H., Yasuda, N., Ichikawa, S.-I., {et~al.} 2002, in Astronomical Data Analysis Software and Systems XI, Vol. 281, 298

\bibitem[{Bartolini(2024)}]{bartolini2024fermi}
Bartolini, C. 2024, The Astronomer's Telegram, 16497, 1

\bibitem[{Basu(1973)}]{basu1973optical}
Basu, D. 1973, The Observatory, Vol. 93, p. 184-189 (1973), 93, 184

\bibitem[{Blandford \& Rees(1978)}]{BlandfordRees1978}
Blandford, R.~D., \& Rees, M.~J. 1978, in Pittsburgh Conference on BL Lac Objects, ed. A.~N. Wolfe (Pittsburgh: University of Pittsburgh Press), 328

\bibitem[{Chatterjee {et~al.}(2012)Chatterjee, Fossati, Urry, Bailyn, Maraschi, Buxton, Bonning, Isler, \& Coppi}]{chatterjee2012optical}
Chatterjee, R., Fossati, G., Urry, C., {et~al.} 2012, The Astrophysical Journal Letters, 763, L11

\bibitem[{Cohen {et~al.}(2014)Cohen, Romani, Filippenko, Cenko, Lott, Zheng, \& Li}]{cohen2014temporal}
Cohen, D.~P., Romani, R.~W., Filippenko, A.~V., {et~al.} 2014, The Astrophysical Journal, 797, 137

\bibitem[{Dermer \& Schlickeiser(1993)}]{dermer1993model}
Dermer, C.~D., \& Schlickeiser, R. 1993, Astrophysical Journal v. 416, p. 458, 416, 458

\bibitem[{Fanaroff \& Riley(1974)}]{fanaroff1974morphology}
Fanaroff, B.~L., \& Riley, J.~M. 1974, Monthly Notices of the Royal Astronomical Society, 167, 31P

\bibitem[{Fossati {et~al.}(1998)Fossati, Maraschi, Celotti, Comastri, \& Ghisellini}]{fossati1998unifying}
Fossati, G.~a., Maraschi, L., Celotti, A., Comastri, A., \& Ghisellini, G. 1998, Monthly Notices of the Royal Astronomical Society, 299, 433

\bibitem[{Ghisellini \& Tavecchio(2008)}]{ghisellini2008blazar}
Ghisellini, G., \& Tavecchio, F. 2008, Monthly Notices of the Royal Astronomical Society, 387, 1669

\bibitem[{Ghisellini \& Tavecchio(2009)}]{ghisellini2009canonical}
---. 2009, Monthly Notices of the Royal Astronomical Society, 397, 985

\bibitem[{Ghisellini {et~al.}(2011{\natexlab{a}})Ghisellini, Tavecchio, Foschini, \& Ghirlanda}]{ghisellini2011transition}
Ghisellini, G., Tavecchio, F., Foschini, L., \& Ghirlanda, G. 2011{\natexlab{a}}, Monthly Notices of the Royal Astronomical Society, 414, 2674

\bibitem[{Ghisellini {et~al.}(2011{\natexlab{b}})Ghisellini, Tagliaferri, Foschini, Ghirlanda, Tavecchio, Ceca, Haardt, Volonteri, \& Gehrels}]{ghisellini2011high}
Ghisellini, G., Tagliaferri, G., Foschini, L., {et~al.} 2011{\natexlab{b}}, Monthly Notices of the Royal Astronomical Society, 411, 901

\bibitem[{Giommi {et~al.}(2012)Giommi, Padovani, Polenta, Turriziani, D’Elia, \& Piranomonte}]{giommi2012simplified}
Giommi, P., Padovani, P., Polenta, G., {et~al.} 2012, Monthly Notices of the Royal Astronomical Society, 420, 2899

\bibitem[{Graham {et~al.}(2020)Graham, Ross, Stern, Drake, McKernan, Ford, Djorgovski, Mahabal, Glikman, Larson, {et~al.}}]{graham2020understanding}
Graham, M.~J., Ross, N.~P., Stern, D., {et~al.} 2020, Monthly Notices of the Royal Astronomical Society, 491, 4925

\bibitem[{Hoffmeister(1929)}]{hoffmeister1929354}
Hoffmeister, C. 1929, Astronomische Nachrichten, volume 236, p. 233, 236, 233

\bibitem[{Kang {et~al.}(2024)Kang, Lyu, Wu, Zheng, \& Fan}]{kang2024physical}
Kang, S.-J., Lyu, B., Wu, Q., Zheng, Y.-G., \& Fan, J. 2024, The Astrophysical Journal, 962, 122

\bibitem[{Kotani {et~al.}(2007)Kotani, Kawai, Yanagisawa, Watanabe, Arimoto, Fukushima, Hattori, Inata, Izumiura, Kataoka, {et~al.}}]{kotani2007mitsume}
Kotani, T., Kawai, N., Yanagisawa, K., {et~al.} 2007, arXiv preprint astro-ph/0702708

\bibitem[{Krawczynski {et~al.}(2002)Krawczynski, Coppi, \& Aharonian}]{krawczynski2002time}
Krawczynski, H., Coppi, P.~S., \& Aharonian, F. 2002, Monthly Notices of the Royal Astronomical Society, 336, 721

\bibitem[{Marscher \& Gear(1985)}]{marscher1985models}
Marscher, A.~P., \& Gear, W.~K. 1985, Astrophysical Journal, Part 1 (ISSN 0004-637X), vol. 298, Nov. 1, 1985, p. 114-127., 298, 114

\bibitem[{Matsubayashi {et~al.}(2019)Matsubayashi, Ohta, Iwamuro, Iwata, Kambe, Tsutsui, Izumiura, Yoshida, \& Hattori}]{matsubayashi2019kools}
Matsubayashi, K., Ohta, K., Iwamuro, F., {et~al.} 2019, Publications of the Astronomical Society of Japan, 71, 102

\bibitem[{Miller {et~al.}(1978)Miller, French, \& Hawley}]{miller1978spectrum}
Miller, J.~S., French, H.~B., \& Hawley, S.~A. 1978, Astrophysical Journal, Part 2-Letters to the Editor, vol. 219, Jan. 15, 1978, p. L85-L87., 219, L85

\bibitem[{Mimica {et~al.}(2009)Mimica, Aloy, Agudo, Mart{\'\i}, G{\'o}mez, \& Miralles}]{mimica2009spectral}
Mimica, P., Aloy, M.-A., Agudo, I., {et~al.} 2009, The Astrophysical Journal, 696, 1142

\bibitem[{Nesci {et~al.}(2021)Nesci, Cutini, Stanghellini, Martinelli, Maselli, Lipunov, Kornilov, Lopez, Siviero, Giroletti, {et~al.}}]{nesci2021multiwavelength}
Nesci, R., Cutini, S., Stanghellini, C., {et~al.} 2021, Monthly Notices of the Royal Astronomical Society, 502, 6177

\bibitem[{Pandey {et~al.}(2024{\natexlab{a}})Pandey, Kushwaha, Wiita, Prince, Czerny, \& Stalin}]{pandey2024origin}
Pandey, A., Kushwaha, P., Wiita, P.~J., {et~al.} 2024{\natexlab{a}}, Astronomy \& Astrophysics, 681, A116

\bibitem[{Pandey {et~al.}(2024{\natexlab{b}})Pandey, Hu, Wang, Czerny, Chen, Songsheng, Wang, Zhang, \& Aceituno}]{pandey2024b2}
Pandey, A., Hu, C., Wang, J.-M., {et~al.} 2024{\natexlab{b}}, The Astrophysical Journal, 978, 120

\bibitem[{Pe{\~n}a-Herazo {et~al.}(2021)Pe{\~n}a-Herazo, Massaro, Gu, Paggi, Landoni, D’Abrusco, Ricci, Masetti, \& Chavushyan}]{pena2021optical}
Pe{\~n}a-Herazo, H.~A., Massaro, F., Gu, M., {et~al.} 2021, The Astronomical Journal, 161, 196

\bibitem[{Rajput {et~al.}(2019)Rajput, Stalin, Sahayanathan, Rakshit, \& Mandal}]{rajput2019temporal}
Rajput, B., Stalin, C., Sahayanathan, S., Rakshit, S., \& Mandal, A.~K. 2019, Monthly Notices of the Royal Astronomical Society, 486, 1781

\bibitem[{Ricci \& Trakhtenbrot(2023)}]{ricci2023changing}
Ricci, C., \& Trakhtenbrot, B. 2023, Nature Astronomy, 7, 1282

\bibitem[{Ruan {et~al.}(2014)Ruan, Anderson, Plotkin, Brandt, Burnett, Myers, \& Schneider}]{ruan2014nature}
Ruan, J., Anderson, S., Plotkin, R., {et~al.} 2014, The Astrophysical Journal, 797, 19

\bibitem[{Sako {et~al.}(2016)Sako, Osawa, Takahashi, Kikuchi, Doi, Kobayashi, Aoki, Arimatsu, Ichiki, Ikeda, {et~al.}}]{sako2016development}
Sako, S., Osawa, R., Takahashi, H., {et~al.} 2016, in Ground-based and Airborne Instrumentation for Astronomy VI, Vol. 9908, International Society for Optics and Photonics, 99083P

\bibitem[{Schmidt(1963)}]{schmidt19633c273}
Schmidt, M. 1963, A Century of Nature, Twenty One Discoveries That Changed Science and the World

\bibitem[{Shimokawabe {et~al.}(2008)Shimokawabe, Kawai, Kotani, Yatsu, Ishimura, Vasquez, Mori, Kudo, Yoshida, Yanagisawa, {et~al.}}]{shimokawabe2008mitsume}
Shimokawabe, T., Kawai, N., Kotani, T., {et~al.} 2008, AIP Conference Proceedings, 1000, 543

\bibitem[{Sikora(1994)}]{sikora1994high}
Sikora, M. 1994, in International Astronomical Union Colloquium, Vol. 142, Cambridge University Press, 923--928

\bibitem[{Takase {et~al.}(1977)Takase, Ishida, Shimizu, Maehara, Hamajima, Noguchi, \& Ohashi}]{takase1977105}
Takase, B., Ishida, K., Shimizu, M., {et~al.} 1977, Annals of the Tokyo Astronomical Observatory, 16, 74

\bibitem[{Tchekhovskoy {et~al.}(2011)Tchekhovskoy, Narayan, \& McKinney}]{tchekhovskoy2011efficient}
Tchekhovskoy, A., Narayan, R., \& McKinney, J.~C. 2011, Monthly Notices of the Royal Astronomical Society: Letters, 418, L79

\bibitem[{Urry \& Padovani(1995)}]{urry1995unified}
Urry, C.~M., \& Padovani, P. 1995, Publications of the Astronomical Society of the Pacific, 107, 803

\bibitem[{Vestergaard \& Wilkes(2001)}]{vestergaard2001empirical}
Vestergaard, M., \& Wilkes, B.~J. 2001, The Astrophysical Journal Supplement Series, 134, 1

\bibitem[{Wald(1974)}]{wald1974black}
Wald, R.~M. 1974, Physical Review D, 10, 1680

\bibitem[{Yanagisawa {et~al.}(2010)Yanagisawa, Kuroda, Yoshida, Shimizu, Nagayama, Toda, Ohta, \& Kawai}]{yanagisawa2010six}
Yanagisawa, K., Kuroda, D., Yoshida, M., {et~al.} 2010, AIP Conference Proceedings, 1279, 466

\bibitem[{Yatsu {et~al.}(2007)Yatsu, Kawai, Shimokawabe, Vasquez, Ishimura, Kotani, Yanagisawa, Yoshida, Nagayama, Shimizu, {et~al.}}]{yatsu2007development}
Yatsu, Y., Kawai, N., Shimokawabe, T., {et~al.} 2007, Physica E: Low-dimensional Systems and Nanostructures, 40, 434

\bibitem[{York {et~al.}(2000)York, Adelman, Anderson~Jr, Anderson, Annis, Bahcall, Bakken, Barkhouser, Bastian, Berman, {et~al.}}]{york2000sloan}
York, D.~G., Adelman, J., Anderson~Jr, J.~E., {et~al.} 2000, The Astronomical Journal, 120, 1579

\bibitem[{Zhang {et~al.}(2024)Zhang, Doi, Kokubo, Sako, Ohsawa, Tominaga, Tanaka, Fukazawa, Takahashi, Arima, {et~al.}}]{zhang2024optical}
Zhang, T., Doi, M., Kokubo, M., {et~al.} 2024, The Astrophysical Journal, 968, 71

\end{thebibliography}
\bibliographystyle{aasjournal}



\end{document}